\documentclass[10pt,letterpaper]{article}
\usepackage[top=0.85in,left=0.85in,footskip=0.75in]{geometry}
\usepackage{amsmath,amssymb}

\usepackage{changepage}

\usepackage[utf8x]{inputenc}

\usepackage{textcomp,marvosym}

\usepackage{cite}

\usepackage{nameref,hyperref}

\usepackage{microtype}
\DisableLigatures[f]{encoding = *, family = * }

\usepackage[table]{xcolor}

\usepackage{array}

\newcolumntype{+}{!{\vrule width 2pt}}

\newlength\savedwidth

\newcommand\thickhline{\noalign{\global\savedwidth\arrayrulewidth\global\arrayrulewidth 2pt}%
	\hline
	\noalign{\global\arrayrulewidth\savedwidth}}

\usepackage[aboveskip=1pt,labelfont=bf,labelsep=period,justification=raggedright,singlelinecheck=off]{caption}

\makeatletter
\renewcommand{\@biblabel}[1]{\quad#1.}
\makeatother

\date{}

\usepackage{lastpage,fancyhdr,graphicx}
\usepackage{epstopdf}
\usepackage[final]{pdfpages}

\begin{document}

\vspace*{0.2in}

\begin{flushleft}
{\Large
\textbf\newline{Feedback Between Motion and Sensation Provides Nonlinear Boost in Run-and-tumble Navigation} % Please use "title case" (capitalize all terms in the title except conjunctions, prepositions, and articles).
}
\newline
\\
J.~Long\textsuperscript{1,2},
S.W.~Zucker\textsuperscript{3,4},
T.~Emonet\textsuperscript{2,1*}
\\
\bigskip
\textbf{1} Department of Physics, Yale University, New Haven, CT, USA
\\
\textbf{2} Department of Molecular, Cellular and Developmental Biology, Yale University, New Haven, CT, USA
\\
\textbf{3} Department of Computer Science, Yale University, New Haven, CT 06520, USA
\\
\textbf{4} Department of Biomedical Engineering, Yale University, New Haven, CT 06520, USA
\\
\bigskip
% Use the asterisk to denote corresponding authorship and provide email address in note below.
* Corresponding author
E-mail: thierry.emonet@yale.edu (TE)

\end{flushleft}

\section*{Abstract}
Many organisms navigate gradients by alternating straight motions (runs) with random reorientations (tumbles), transiently suppressing tumbles whenever attractant signal increases. This induces a functional coupling between movement and sensation, since tumbling probability is controlled by the internal state of the organism which, in turn, depends on previous signal levels. Although a negative feedback tends to maintain this internal state close to adapted levels, positive feedback can arise when motion up the gradient reduces tumbling probability, further boosting drift up the gradient. Importantly, such positive feedback can drive large fluctuations in the internal state, complicating analytical approaches. Previous studies focused on what happens when the negative feedback dominates the dynamics. By contrast, we show here that there is a large portion of physiologically-relevant parameter space where the positive feedback can dominate, even when gradients are relatively shallow. We demonstrate how large transients emerge because of non-normal dynamics \textcolor{black}{(non-orthogonal eigenvectors near a stable fixed point)} inherent in the positive feedback, and further identify a fundamental nonlinearity that strongly amplifies their effect. Most importantly, this amplification is asymmetric, elongating runs in favorable directions and abbreviating others.  The result is a ``ratchet-like'' gradient climbing behavior with drift speeds that can approach half the maximum run speed of the organism. Our results thus show that the classical drawback of run-and-tumble navigation --- wasteful runs in the wrong direction --- can be mitigated by exploiting the non-normal dynamics implicit in the run-and-tumble strategy.
%Countless organisms use a run-and-tumble strategy to navigate gradients. The classical drawback of this approach is that runs in the wrong direction are wasteful. We show analytically that organisms can overcome this fundamental limitation by exploiting the non-normal dynamics and intrinsic nonlinearities inherent to the positive feedback between motion and sensation. This mechanism drives large asymmetric fluctuations in the organism's internal state, not described by mean field theory, that selectively amplify runs in the correct direction and result in fast ``ratchet-like'' gradient climbing.

% Please keep the Author Summary between 150 and 200 words
% Use first person. PLOS ONE authors please skip this step. 
% Author Summary not valid for PLOS ONE submissions.   
\section*{Author Summary}
Countless bacteria, larvae and even larger organisms (and robots) navigate gradients by alternating periods of straight motion (runs) with random reorientation events (tumbles). Control of the tumble probability is based on previously-encountered signals. A drawback of this run-and-tumble strategy is that occasional runs in the wrong direction are wasteful. Here we show that there is an operating regime within the organism’s internal parameter space where run-and-tumble navigation can be extremely efficient. We characterize how the positive feedback between behavior and sensed signal results in a type of non-equilibrium dynamics, with the organism rapidly tumbling after moving in the wrong direction and extending motion in the right ones. For a distant source, then, the organism can find \textcolor{black}{it} fast.

%\linenumbers

\section*{Introduction}

Navigation up a gradient succeeds by finding those directions in which signals of interest increase. This can be difficult when the size of the navigator is small compared to the length scale of the gradient because local directional information becomes unreliable. In this case, cells~\cite{bergbrown,marineMicrobes}, worms~\cite{albrecht-bargmann}, larvae~\cite{drosophilaChemotaxis}, and even robots~\cite{swarmRobotics} often adopt a run-and-tumble strategy to navigate. During runs the organism moves approximately straight, collecting differential sensor information in one direction. Tumbles, or reorientation\textcolor{black}{s} at zero speed, enable the organism to explore other directions. Signal levels are transduced rapidly to the motility apparatus through an internal state variable, so that increases in attractant transiently raise the probability to run longer (and tumble less) before a negative integral feedback adapts it back~\cite{doyle}. Classically, averaging over many runs and tumbles results in a net drift up the gradient, although this is usually rather modest because of \textcolor{black}{the} occasional runs in the wrong direction. We here focus on the positive feedback inherent to this strategy wherein motion up the gradient lowers the probability to tumble, which further boosts drift up the gradient. Our analysis reveals an unstudied regime in which rapid progress can be achieved. \textcolor{black}{Small fluctuations in the speed of the organism along the gradient grow into large transients in the correct direction but small ones otherwise. We show that this asymmetric amplification arises from the positive feedback, which causes the eigenvectors near the adapted state of the dynamical system to become non-orthogonal, therefore leading to non-normal dynamics. The resulting large transient are further boosted by a nonlinearity that is intrinsic to the positive feedback. Such non-normal dynamics were first discovered in fluid mechanics where they were shown to play an important role in the onset of turbulence in the absence of unstable modes~\cite{trefethen,nonmodal}}.

Past theoretical studies of run-and-tumble navigation have mostly focused on what happens when adaptation dominates the dynamics (e.g.~\cite{KS,schnitzer,vergassola,tuyuhai,yann}). In this regime, the internal state of the organism exhibits small fluctuations around its mean, and mean field theory (MFT) can be applied to make predictions. This approach has been used to describe the motile behavior or populations of \textit{E. coli} bacteria in exponential ramps~\cite{shimizu-tu-berg,nick,yann} and oscillating gradients~\cite{frequencyDependent}. Beyond the well-understood negative feedback-dominated regime there is a large portion of physiologically relevant parameter space where the positive feedback between movement and sensation dominates the run-and-tumble dynamics. Agent-based simulations have shown that\textcolor{black}{,} in this case\textcolor{black}{,} large transient fluctuations can emerge in the internal state of an individual organism climbing a gradient, precluding the use of mean field approaches~\cite{yann}. While systems of partial differential equations (PDEs) can be integrated numerically to reproduce these dynamics~\cite{xue-yang}, a precise understanding of the role of the positive feedback in generating such large fluctuations and the impact of those on the performance of a biased random walk are fundamental questions that remain largely unanswered because of difficulties in obtaining analytical results.

Here we develop an analytical model of run-and-tumble gradient ascent that preserves the rich nonlinearity of the problem and incorporates the internal state, 3D-direction of motion, and position of the organism as stochastic variables. We find that large fluctuations in the internal state originate from two key mechanisms: (i) the non-normal dynamical structure of the positive feedback that enables small fluctuations to grow, and (ii) a quadratic nonlinearity in the speed along \textcolor{black}{the} gradient that further amplifies such transients asymmetrically. \textcolor{black}{Utilizing} phase space analysis and stochastic simulations, we show how these two effects combine to generate a highly effective ``ratchet-like'' gradient-climbing mode that strongly mitigates the classic drawback of biased random walks: wasteful runs in wrong directions. In this \textcolor{black}{new} regime an organism should be able to achieve drift speeds on the order of the maximum swim speed. Our results are general in that they apply to a large class of biased random walk strategies\textcolor{black}{,} where run speed and sampling of new directions may be modulated based on \textcolor{black}{previously encountered signals}.

\section*{Results}

\subsection*{Minimal model of run-and-tumble navigation}

Consider a random walker with an internal state variable $F$ that follows linear relaxation towards the adapted state $F_0$ over the timescale $t_M$, which represents the memory duration of the random walker. We assume that the \textcolor{black}{perceived} signal\textcolor{black}{,} $\phi({\bf X}, t)\textcolor{black}{=\phi(C({\bf X}, t))}$\textcolor{black}{,} at position ${\bf X}$ and time $t$ \textcolor{black}{(here $C$ represents the signal),} is rapidly transduced to determine the value of an internal state variable $F$ via a receptor with gain $N$:
\begin{equation} \label{eq:dFdt}
	\dot{F}= - \frac{F-F_0}{t_M} + N \left(\frac{\partial}{\partial t}+\dot{{\bf X}} \cdot \nabla \right) \phi(\bf X, t).
\end{equation}
Stochastic switching between runs and tumbles depends on $F$ and follows inhomogeneous Poisson statistics with probability to run $r(F) = \lambda_T / \left(\lambda_R + \lambda_T\right) = 1/\left(1+\exp\left(- H F\right)\right)$, where $H$ is the gain of the motor, and $\lambda_R(F)$ and $\lambda_T(F)$ are the transition rates from run to tumble and vice versa~\cite{tuReview, nick}.

During runs the speed is constant $\|\dot{\bf{X}}\|=v_0$ and the direction of motion is subject to rotational Brownian motion with diffusion coefficient $D_R$. During tumbles the speed is nil and reorientation follows rotational diffusion $D_T>D_R$ to account for persistence \textcolor{black}{effects}~\cite{saragosti2012}. Taken together\textcolor{black}{,} these two processes cause the random walker to lose its original direction at the expected rate $t_D^{-1} = (n-1)(r D_R + (1-r D_T)$ where $n=2,3$ for two- and three-dimensional motion respectively. \textcolor{black}{Note that, in this minimal model we ignore possible internal signaling noise~\cite{michael,tenwolde}, and all randomness comes from the rotational diffusions $D_R$ and $D_T$ as well as the stochastic switchings with rates $\lambda_R(f)$ and $\lambda_T(f)$. The effect of signaling noise is considered below using agent-based simulations. Since $\phi(C)$ can be nonlinear, Eq~\eqref{eq:dFdt} includes possible effects of saturation of the sensory system}.

Consider \textcolor{black}{a} static one-dimensional gradient \textcolor{black}{and define the length scale of the perceived gradient as} $L({\bf X})= 1 /||\nabla \phi({\bf X})||$ \textcolor{black}{and the direction of motion as $s= \hat{\bf{u}} \cdot \hat{\bf{X}}$. Then from Eq~\eqref{eq:dFdt} the internal dynamics satisfies the following equations during runs and tumbles, respectively:
\begin{equation} \label{eq:dFdtrt}
	\begin{aligned}
		&\textcolor{black}{\dot{F}|_{run}} &\textcolor{black}{=} &\textcolor{black}{- \frac{F-F_0}{t_M} + \frac{N v_0}{L(\bf X)} s}  \\
		&\textcolor{black}{\dot{F}|_{tumble}} &\textcolor{black}{=} &\textcolor{black}{- \frac{F-F_0}{t_M}}.
	\end{aligned}
\end{equation}
}
We are interested in the displacement of the random walker along the gradient over timescales longer than individual runs and tumbles. In the limit where the switching timescale $t_S= 1/(\lambda_R + \lambda_T)$ is much shorter than the other timescales we \textcolor{black}{derive from a two-state stochastic model and Eq~\eqref{eq:dFdtrt}} (Methods Eqs~\eqref{eq:master-rescale}--\eqref{eq:FP-prescale}):
\begin{equation} \label{eq:drdt}
\dot{r} = r\left(1-r\right) \bigg( \underbrace{-\frac{f(r)-f_0}{t_M} }_{\text{negative feedback}} + \underbrace{\frac{r\,s}{L\textcolor{black}{(\bf X)}/(N H v_0)}}_{\text{positive feedback}} \bigg),
\end{equation}
where $f = H F$. The first term is the negative feedback towards the adapted run probability $r_0 = r(f_0)$. The second term shows how motion up the gradient ($s>0$) causes the probability to run $r$ to feed back on itself \textcolor{black}{--- when the organism is oriented up the gradient ($s>0$), $F$ increases only during runs (Eq~\eqref{eq:dFdtrt}), and this increase in turn raises $r(F)$ so that the probability that the dynamics of $F$ follows $\dot{F}|_{run}$ rather than $\dot{F}|_{tumble}$ is increased, and so on}. \textcolor{black}{A} positive feedback \textcolor{black}{is thereby created} with characteristic timescale $t_E=L/(N H v_0)$\textcolor{black}{. Steeper gradient (smaller $L$), stronger receptor gain $N$ or motor gain $H$, or faster speed $v_0$, all lead to stronger positive feedback (shorter $t_E$). This important timescale, $t_E$,} together with $t_M$ (memory duration) and $t_D$ (direction decorrelation time)\textcolor{black}{, effectively} determines the dynamics.

Expressing time in units of $t_M$\textcolor{black}{,} we \textcolor{black}{introduce} the following two non-dimensional timescales:
\begin{equation} \label{eq:timescales}
\begin{aligned}
\tau_E\textcolor{black}{({\bf X})}   &=\frac{t_E\textcolor{black}{({\bf X})}}{t_M} = \frac{L\textcolor{black}{({\bf X})}}{t_M N H v_0} \\
\tau_D\textcolor{black}{(f)}   &=\frac{t_D}{t_M} = \frac{1/t_M}{(n-1) \left(r\textcolor{black}{(f)} D_R+\left(1-r\textcolor{black}{(f)}\right) D_T\right)}.
\end{aligned}
\end{equation}
Here $\tau_E$ quantifies the ratio between the negative and positive feedbacks. \textcolor{black}{(See Table~\ref{tab:symbols} for a summary of the symbols used.)} \textcolor{black}{From above, w}e expect that the dynamics will depend on how $\tau_E$ and $\tau_D$ compare with one.

\subsection*{Exploration of the dynamical regimes}

\begin{figure}[!h]
		\includegraphics[width=\textwidth]{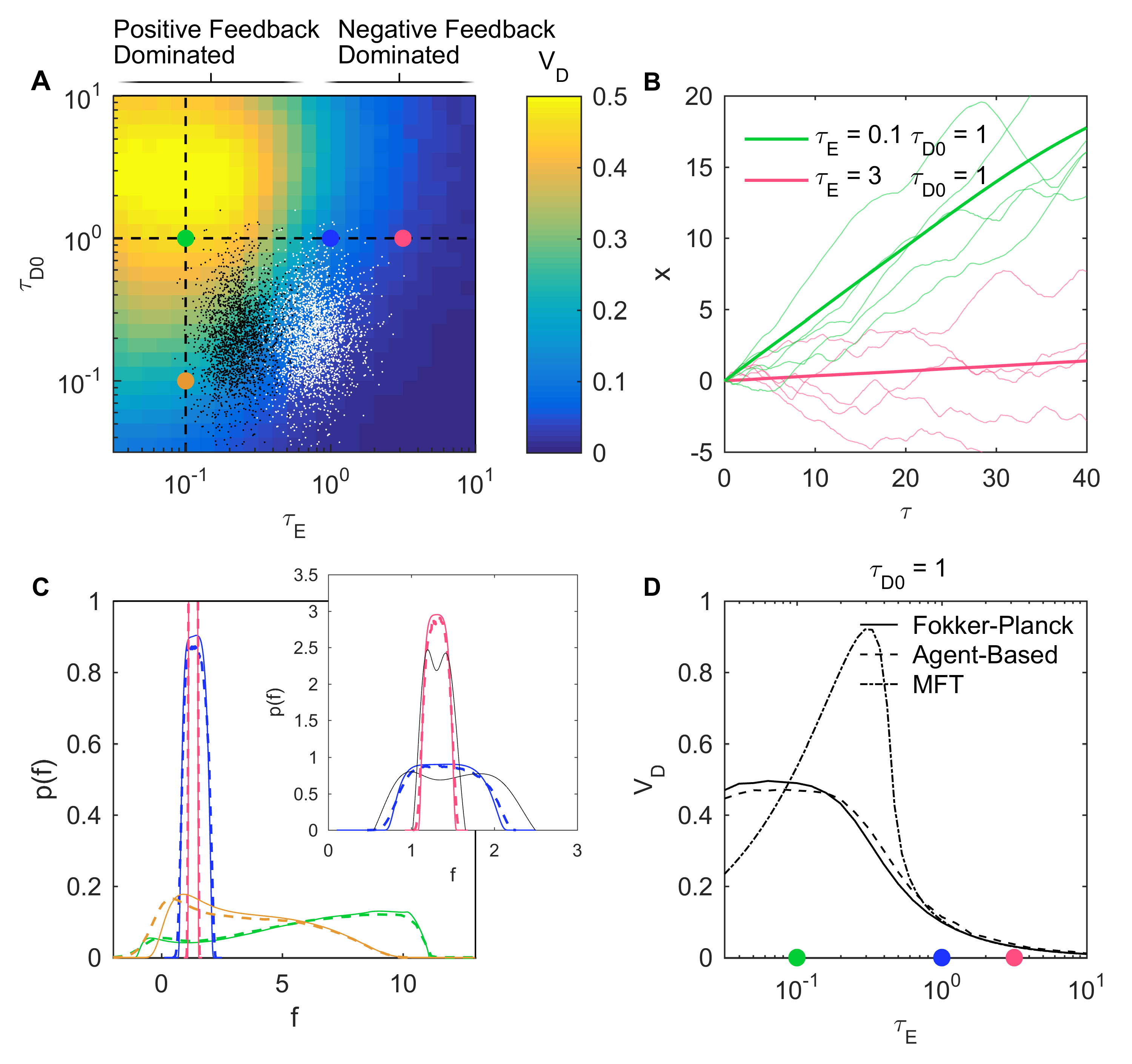}
	\caption{{\bf Different dynamical regimes of run-and-tumble gradient ascent.}
		(A) Drift speed $V_D$ of simulated \textit{E. coli} cells swimming in static exponential gradients as a function of $\tau_E$ and $\tau_{D0}$. Green, blue, and red: $\tau_{D0}=1$ and $\tau_E=0.1, 1, 3$, respectively. Orange: $\tau_E=0.1$ and $\tau_{D0}=0.1$ (dashed line: guides to the eye). White/black: sampling of a wild type population~\cite{adam} near the bottom/top of a linear gradient. (B) Classical (red) vs. rapid climbing (green) trajectories. $x=X/(v_0 t_M)$ vs. time $\tau=t/t_M$ for cells in the positive-feedback- (green) and negative-feedback-dominated (red) regime (thin: 5 samples; thick: mean over $10^4$ samples). (C) Marginal probability distribution of the internal variable at steady state $\overline{p}(f)$; solid: numerical solution of Eq~\eqref{eq:FP}; dashed: sampled distribution from agent-based simulation; colors: same parameter values as in A. Inset: zoomed view with second order analytical approximations (Methods Eq~\eqref{eq:pexp}) in black. $r_0=0.8$ and $D_T/D_R=37$ in all simulations. (D) Comparison of different methods to calculate $V_D$ as a function of $\tau_E$ keeping $\tau_{D0} = 1$ fixed. Solid: numerical integration of Eq~\eqref{eq:FP} and Eq~\eqref{eq:VD}; dashed: agent-based model simulations; dash-dot: MFT (Methods Eq~\eqref{eq:VDMFT_m}). Details in Methods.}
	\label{fig:Fig1}
\end{figure}

To explore how run-and-tumble dynamics depend on $\tau_E$ and $\tau_D$, we used a previously published stochastic agent-based simulator of the bacteria \textit{E. coli} that reproduces well available experimental data on the wild-type laboratory strain RP437 (~\cite{nick,adam} and \nameref{S1}). In this case the internal state $F$ represents the free energy of the chemoreceptors. Since \textit{E. coli} approximately detects log-concentrations (\nameref{S1} Eq~(S11)), we simulated an exponential gradient so that $\tau_E$ is a constant. In this case the cells reach steady state with a constant drift speed $V_D$. Calculating $V_D$ from $10^4$ simulated trajectories for a range of $\tau_E$ and $\tau_{D0}=\tau_D(r_0)$ values reveals that cells climb the gradient much faster when the positive feedback dominates ($\tau_E < 1$) (Fig~\ref{fig:Fig1}A). The trajectories of individual cells resembled that of a ratchet that moves almost only in one direction (Fig~\ref{fig:Fig1}B green). In contrast, when the negative feedback dominates ($\tau_E > 1$) the trajectories exhibit both up and down runs of similar \textcolor{black}{although} slightly biased lengths (Fig~\ref{fig:Fig1}B red). $V_D$ also depends on $\tau_D$ and peaks when the direction decorrelation time is on the same order as the memory duration ($\tau_D\simeq 1$), consistent with previous studies~\cite{rava,damon,vergassola}. In these simulations the adapted probability to run $r_0=0.8$ and the ratio $D_T/D_R=37$ were kept constant. Changing these values did not change the main results (\nameref{S1_Fig}(A-C)).

In a wild type population, individual isogenic cells will have different values of $\tau_E$ and $\tau_{D0}$ due to cell-to-cell variabilities in swimming speed and in the abundance of chemotaxis proteins~\cite{yannExperimental,adam,spudichKoshland1976}. \textcolor{black}{In a recent experimental study, the phenotype and performance of individual wild type cells (RP437 strain) was quantified by tracking cells swimming up a static quasi-linear gradient of methyl-aspartate (varying from $0$ to $1~mM$ over $10~mm$). This experiment revealed large differences among the performances of individual cells within the isogenic population~\cite{adam}, which could be reproduced by complementing the model of bacterial chemotaxis just described with a simple model of noisy gene expression (Fig 2 in~\cite{adam}).} To examine in which region of the $(\tau_E,\tau_{D0})$ space \textcolor{black}{these cells might have been operating, we used this same model (complemented with diversity in rotational diffusion coefficients $D_R$ and $D_T$ due to variations in cell length; see \nameref{S1}) with the same parameter values to run simulations of 16,000 cells climbing the experimentally measured (Fig~\ref{fig:Fig1}A)}. We find that even in this relatively shallow gradient some cells \textcolor{black}{might have been operating} in the positive-feedback-dominated regime, especially near the bottom of the gradient (black dots). As the cells climb the gradient\textcolor{black}{,} $\tau_E$ becomes larger \textcolor{black}{(white dots) because\textcolor{black}{,} as concentration increases, the log-sensing cells in the quasi-linear gradient face a shallower gradient, and thus weaker positive feedback}.

\subsection*{Positive feedback between motion and sensation generates large internal state fluctuations and fast drift}

To better understand the origin of the fast drift speed and its associated ``ratchet-like'' behavior, we examine the relationship between the drift speed $V_D$ and the statistics of the internal state $f$. Using $t_M$ as the unit of time and $v_0 t_M$ as the unit of length we derive a Fokker-Planck equation for the probability $P(x, f, s,\tau)$ that at time $\tau = t/t_M$ the cell is at position $x=X/(v_0 t_M)$ with internal state $f$ and orientation $s$ (Methods Eq~\eqref{eq:FP-prescale}):
\begin{equation} \label{eq:FP}
	\begin{aligned}
		\partial_{\tau} P = &- \partial_f \left( \left(-\left(f-f_0\right)+ \frac{r\textcolor{black}{(f)} s}{\tau_E\textcolor{black}{(x)}}\right) P \right) + \frac{\hat{L}_s P}{(n-1)\tau_D\textcolor{black}{(f)}} - \partial_{x} \left( r\textcolor{black}{(f)} s P \right).
	\end{aligned}
\end{equation}
Here $\hat{L}_s = (1-s^2)^{\frac{3-n}{2}} \partial_s ((1-s^2)^{\frac{n-1}{2}} \partial_s )$ is the rotational diffusion operator on the ($n-1$)-sphere. \textcolor{black}{All symbols used are summarized in Table~\ref{tab:symbols}.}

\textcolor{black}{For simplicity we consider a log-sensing organism swimming in a static exponential gradient. In this case, $\tau_E(x) = \tau_E$ is constant (more complex gradient profiles and the effect of receptor saturation are considered later in the paper). Therefore the positive feedback becomes independent of position and the system can reach a steady state drift speed. Separating the variable $x$ and integrating over $x$ we obtain} (Methods Eq~\eqref{eq:VDderive})
\begin{equation} \label{eq:VD}
	\begin{aligned}
		V_D = \tau_E \overline{\langle f-f_0\rangle},
	\end{aligned}
\end{equation}
where $\langle\cdot\rangle$ represents averaging over $f$ and $s$ and the bar indicates steady state. Eq~\eqref{eq:VD} indicates that the drift speed is determined by the steady state marginal distribution $\overline{p}(f)$. To find an analytical expression for $\overline{p}(f)$, we expand the steady state joint distribution $\overline{P}(f,s)$ in orthonormal eigenfunctions of the angular operator $\hat{L}_s$ --- the first two coefficients are the marginal distribution $\overline{p}(f)$ and the first angular moment $\overline{p}_1(f)/\sqrt{n} = \int \overline{P} s \mathrm{d}s$ --- and discard higher orders to obtain a closed system of equations. \textcolor{black}{The analytical solution for the steady state marginal distribution $\overline{p}(f)$ reads 
\begin{equation} \label{eq:dpf_main}
	\begin{aligned}
		\overline{p}(f)=&  \frac{1}{W}\frac{\frac{r(f)}{\tau_E}}{\frac{1}{n} \left(\frac{r(f)}{\tau_E}\right)^2 - \left(f-f_0\right)^2} \exp\left( - \int^f \frac{ \frac{f_1-f_0}{\tau_D(f_1)} }{\frac{1}{n} \left(\frac{r(f_1)}{\tau_E}\right)^2 - \left(f_1-f_0\right)^2} \mathrm{d} f_1 \right),
	\end{aligned}
\end{equation}
where $W$ is a normalization constant. The full derivation is provided in Methods Eqs~\eqref{eq:pkPDEs}--\eqref{eq:dpf}, together with an interpretation of the distribution as a potential solution $\overline{p}(f) \propto \exp -V(f)$ where $V(f)$ is the ``potential''. We also examine how the shape of the potential depends on $\tau_E$ and $\tau_D$.}

The solution $\overline{p}(f)$ is plotted in Fig~\ref{fig:Fig1}C. When the negative feedback dominates ($\tau_E\gtrsim 1$) the distribution is sharply peaked and nearly Gaussian with variance $\sigma^2 = \tau_{D0} r_0^2/n\tau_E^2$ (Methods Eq~\eqref{eq:variance}) and its mean barely deviates from the adapted state $f_0$ (Fig~\ref{fig:Fig1}C red and blue). Substituting $\overline{p}(f)$ into Eq~\eqref{eq:VD} and taking the limit $\tau_E \gg 1$ yields known MFT results~\cite{vergassola,yann,tuyuhai} (Methods Eq~\eqref{eq:VDMFT_m}). When the positive feedback dominates ($\tau_E \ll 1$) the distribution $\overline{p}(f)$ now exhibits large asymmetrical deviations (Fig~\ref{fig:Fig1}C green) between the lower and upper bounds $f_L$ and $f_U$, which satisfy the relations $f_L=f_0-r(f_L)/\tau_E$ and $f_U=f_0+r(f_U)/\tau_E$. For small $\tau_E$ the lower bound decreases as $f_L \rightarrow \ln \tau_E$ whereas the upper bound increases as $f_U \rightarrow 1/\tau_E$ (Methods Eq~\eqref{eq:fBounds}). MFT becomes inadequate in this regime, as recently suggested by 1D approximations~\cite{xue-yang}. When the positive feedback dominates, matching the memory of the cell with the direction decorrelation time becomes important: keeping the direction of motion long enough ($\tau_D \gtrsim 1$) allows the distribution to develop a peak near $f_U$ (Fig~\ref{fig:Fig1}C green), which according to Eq~\eqref{eq:VD} results in higher drift speed~(\nameref{S2_Fig}). We verified the approximate analytical solution $\overline{p}(f)$ captures the run-and-tumble dynamics well by plotting it against the distribution of $f$ obtained from the agent-based simulations (Fig~\ref{fig:Fig1}C). Integrating $\overline{p}(f)$ according to Eq~\eqref{eq:VD} predicts well the drift speed for all $\tau_E$ (Fig~\ref{fig:Fig1}D), including where the positive feedback dominates ($\tau_E <1$).

\subsection*{Nonlinear amplification of non-normal dynamics generate\textcolor{black}{s} long runs uphill but short ones otherwise}

\begin{figure}[!h]
		\includegraphics[width=\textwidth]{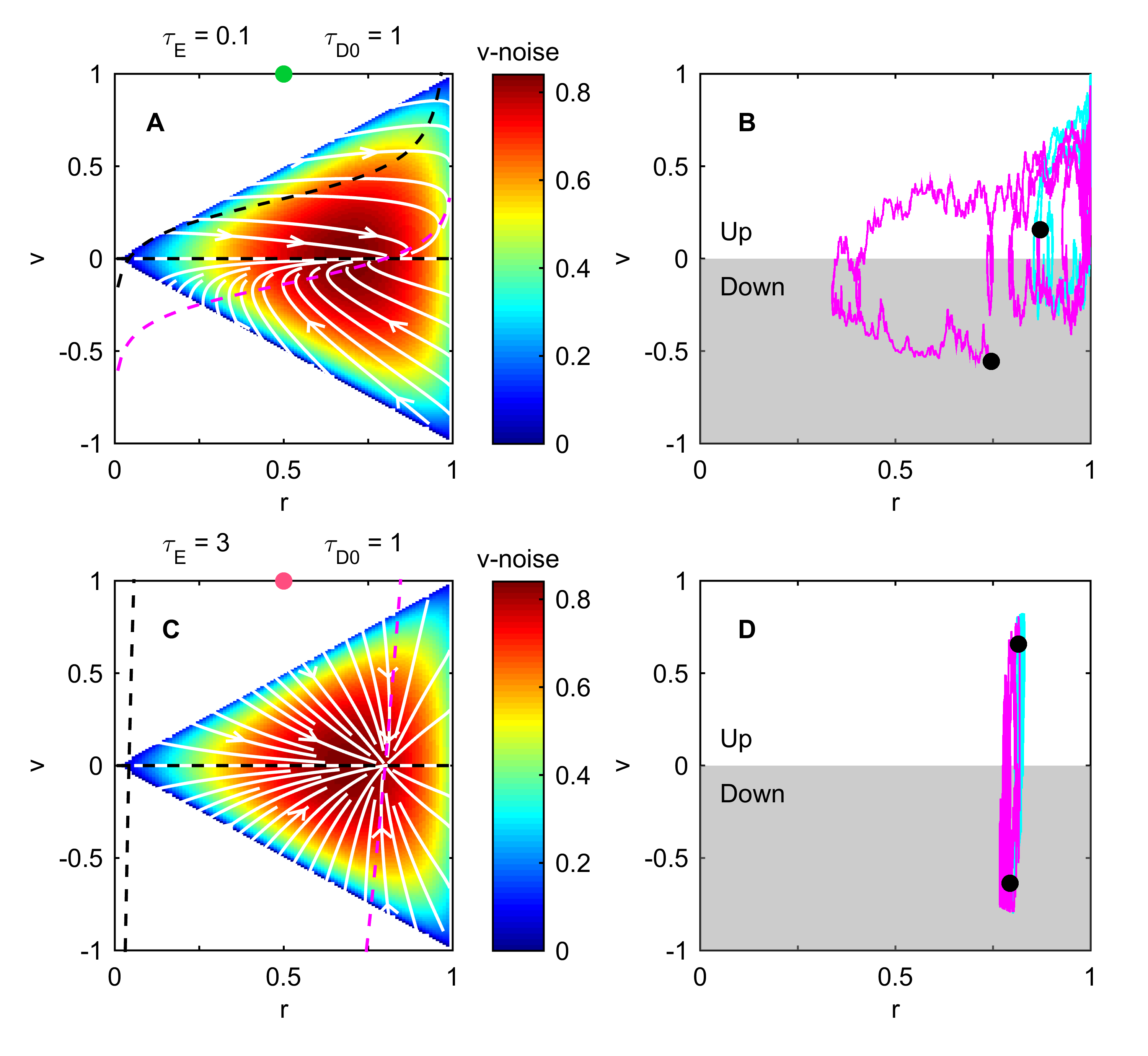}
	\caption{{\bf Non-normal dynamics enables large asymmetric transients in internal state.}
		(A) Phase space diagram of Eq~\eqref{eq:Langevin} when the positive feedback dominates, $\tau_E=0.1$. White: streamlines without noise; magenta: \textcolor{black}{the} $r$-nullcline \textcolor{black}{where $\mathrm{d}r/\mathrm{d}\tau=0$}; black: \textcolor{black}{the} two $v$-nullclines \textcolor{black}{where $\mathrm{d}v/\mathrm{d}\tau=0$}. Heat map: noise magnitude \textcolor{black}{of $\mathrm{d}v/\mathrm{d}\tau$ ($\sqrt{(r^2-v^2)/\tau_D(r)}$ in Eq~\eqref{eq:Langevin})}. (B) \textcolor{black}{Two example trajectories starting in positive (cyan) or negative (magenta) direction. Each trajectory starts from black and lasts over the same time period of $\tau=10$. See also~\nameref{S1_Movie}.} (C,D) Same as A,B \textcolor{black}{except in} the negative-feedback-dominated regime, $\tau_E=3$. \textcolor{black}{When the positive feedback dominates ($\tau_E=0.1$, A), the streamlines (white) are highly asymmetric around the fixed point. They tend to bring the system transiently towards $r=1$ and $v=1$ --- a result of both non-normal dynamics (non-orthogonal eigenvectors near the fixed point) and nonlinear positive feedback (growth towards $v=1$ away from the fixed point) --- before eventually falling back to the fixed point. High noise near the fixed point causes the system to quickly move away from it (magenta in B). Low noise in the upper right corner ($r=1$ and $v=1$) facilitates longer runs in the correct direction (cyan in B). Taken together, these effects result in a fast ``ratchet-like'' gradient climbing behavior. In contrast, when the negative feedback dominates ($\tau_E=0.1$, C) the streamlines all point back directly to the fixed point and small deviations do not grow (cyan and magenta in D).} Details in Methods.}
	\label{fig:Fig2}
\end{figure}

In the fast gradient climbing regime ($\tau_E\ll 1$) trajectories resemble that of a ratchet (Fig~\ref{fig:Fig1}B). To gain mechanistic insight into this striking efficiency we examined the Langevin system equivalent to the Fokker-Planck Eq~\eqref{eq:FP}. Defining $v = r s$ as the normalized run speed projected along the gradient, we \textcolor{black}{change variables from $(f, s)$ to $(r, v)$ and} obtain (Methods Eqs~\eqref{eq:Pangular}-\eqref{eq:f-dyn}):

\begin{equation} \label{eq:Langevin}
	\begin{aligned}
		\frac{\mathrm{d} r }{\mathrm{d}\tau} =& ~r\left(1-r\right)\left( - \left(f(r)-f_0\right) + \frac{v}{\tau_E} \right)\\
		\frac{\mathrm{d} v }{\mathrm{d}\tau} =& ~v\left(1-r\right)\left( - \left(f(r)-f_0\right) + \frac{v}{\tau_E} \right) - \frac{v}{\tau_D(r)}	+ \sqrt{\frac{\left(r^2-v^2\right)}{\tau_D(r)}} \eta(\tau), 
	\end{aligned}
\end{equation}

where $v=\mathrm{d}x/\mathrm{d}\tau$ and $\eta(\tau)$ denotes delta-correlated Gaussian white noise. The nullclines of the system (Fig~\ref{fig:Fig2}A,C) intersect at the only stable fixed point $(r, v)=(r_0, 0)$ of Eq~\eqref{eq:Langevin} where the eigenvalues of the relaxation matrix 
\begin{equation}\label{eq:linear}
	\begin{bmatrix}
		-1 & \left(1-r_0\right)r_0/\tau_E \\
		0  & -1/\tau_{D0}
	\end{bmatrix}
\end{equation}
are both negative (Methods Eq~\eqref{eq:LangevinEigen}). \textcolor{black}{Stochastic fluctuations} due to rotational diffusions $D_R$ and $D_T$ (heat maps in Fig~\ref{fig:Fig2}A,C) continuously push the system away from the fixed point. \textcolor{black}{The magnitude of these fluctuations is large near the fixed point, causing the system to quickly move away. Fluctuations are smaller near $r = 1$ and $v = 1$, enabling the organism to climb the gradient at high speed for a longer time}. Net drift results from spending more time in the region where $v>0$.

Stochastic excursions in the $(r,v)$-plane away from the fixed point exhibit distinctive trajectories depending on the value of $\tau_E$. When the positive feedback dominates ($\tau_E \ll 1$; Fig~\ref{fig:Fig2}A) the eigenvectors of the relaxation matrix, $(1,0)^T$ and $(\frac{\left(1-r_0\right)r_0}{\tau_E}\frac{\tau_{D0}}{\tau_{D0}-1},1)^T$, are highly non-orthogonal. This \textcolor{black}{defines a non-normal dynamics that} enables linear deviations to grow transiently~\cite{trefethen,nonmodal} to feed the nonlinear positive feedback ($v^2$ term second line in Eq~\eqref{eq:Langevin}) leading to large deviations. Importantly, this only happens for runs that start in the correct direction. If the run is in the wrong direction the linear deviation does not grow (Fig~\ref{fig:Fig2}B; see also~\nameref{S1_Movie}). Asymmetry arises because the $v^2$ term is always positive. Similar selective amplification properties are observed in neuronal networks, where non-normal dynamics enables the network to respond to certain signals while ignoring others (including noise)~\cite{neuron,NNampli}. Thus, a random walker running in the correct direction is aided by the positive feedback, which pushes its internal dynamics towards the upper right corner of the phase plane where $r=1$ and $v=1$. If, instead, the run is in the wrong direction ($v < 0$), the nonlinearity pushes the system back into the high noise region near the fixed point where it will rapidly pick a new direction (Fig~\ref{fig:Fig2}B).

In contrast, when the negative feedback dominates ($\tau_E \gtrsim 1$; Fig~\ref{fig:Fig2}C), the eigenvectors become nearly orthogonal. Linear deviations from the fixed point simply relax to the fixed point regardless of the initial direction of the run. Thus runs up and down the gradient are only marginally different in length, resulting in \textcolor{black}{a} small net drift (Fig~\ref{fig:Fig2}D). This key difference between the positive and negative feedback regimes is reflected in the flow field (white curves in Fig~\ref{fig:Fig2}A,C).

\subsection*{Receptor saturation, varying gradient length scales, and trade-offs}

\begin{figure}[!hp]
		\includegraphics[width=\textwidth]{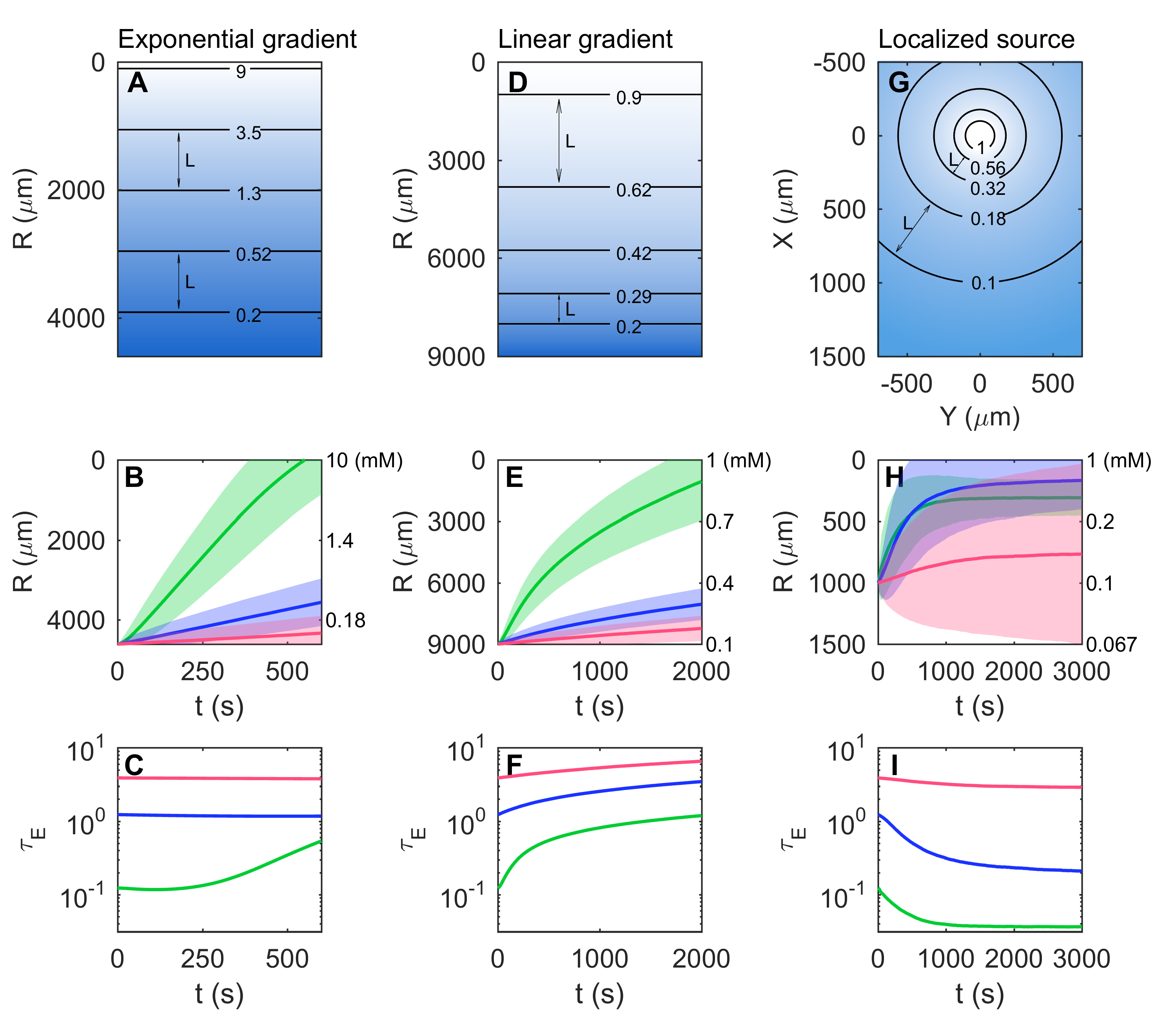}\\
	\caption{{\bf Environmental context, length scales, and receptor saturation.} (A-C) Exponential gradient. (A) Schematic of a gradient of methyl-aspartate $C=C_0 \exp\left(-R/L\textcolor{black}{_0}\right)$ with length scale $L\textcolor{black}{_0}=1000~\mu m$ and source concentration $C_0 = 10~mM$. Contour lines show logarithmically spaced concentration levels in units of $mM$. Contour spacing illustrates \textcolor{black}{constant} $L\textcolor{black}{= 1/|\partial_R \ln C| = L_0}$. (B) The mean trajectory over $10^4$ \textit{E. coli} cells of the position $X$ (in real units $\mu m$) as a function of time $t$ (in $s$) when receptor saturation is taken into account. Initial values of $\tau_E$ are $0.1$ (green), $1$ (blue) or $3$ (red). The shadings indicate standard deviations. The labels on the right axis show the concentration in $mM$ at each position. The effect of receptor saturation becomes visible around $X=4000~\mu m$ (dashed lines) where the concentration of methyl-aspartate is $5.5~mM$. (C) Corresponding time trajectories of the values of $\tau_E$ at mean positions. (D-F) Linear gradient. Similar to A-C but for $C=C_1-a\textcolor{black}{_1}R$ where the source concentration is $C_1 = 1~mM$ and decreases linearly at rate $a\textcolor{black}{_1} = 0.0001~mM/\mu m$ with distance $R$ from the source. Contour spacing decreases with distance from the source (at the top), illustrating decreasing $L\textcolor{black}{= 1/|\partial_R \ln C| = C/a_1 = C_1/a_1-R}$. (G-I) Localized source. Similar to A-C but for \textcolor{black}{a constant} source concentration ($C_2 = 1~mM$) within a ball of radius $R_0 = 100~\mu m$ and \textcolor{black}{for $R>R_0$, the concentration is $C=C_2 R_0/R$ (the steady state solution to the standard diffusion equation $\partial_t C = \nabla^2 C$ without decay), decreasing with radial distance as $1/R$ away from the source. Contour spacing increases away from the source (at the origin), illustrating increasing $L\textcolor{black}{= 1/|\partial_R \ln C| = R}$.}}
	\label{fig:Fig3}
\end{figure}

For simplicity in our analytical derivations we assumed the environment was a constant exponential gradient \textcolor{black}{with} concentrations \textcolor{black}{in} the \textcolor{black}{(log-sensing)} sensitivity range of the organism. Here we explore what happens when the organism encounters concentrations beyond its sensitivity range. For wild type \textit{E. coli} the change in the free energy of the chemorecetor cluster due to ligand binding is proportional to $\ln((1+C/K_i)/(1+C/K_a))$ (\nameref{S1} Eq~(S8)). Therefore the receptor is log-sensing to methyl-aspartate only for concentrations between $K_i\ll C\ll K_a$, where $K_i = 0.0182~mM$ and $K_a = 3~mM$ \textcolor{black}{are the dissociation constants of the inactive and active states of the receptor}. When $C<K_i$ the receptor senses linear concentration~\cite{colin}, whereas when $C>K_a$ the receptors saturate~\cite{keymer-endres-wingreen,hansen-endres-wingreen, sourjik-berg,mello-tu}: as a cell approaches a high concentration source its sensitivity decreases (\nameref{S1} Eq~(S10)). This in turn increases the value of $\tau_E$. Simulations in an exponential gradient show that this effect results in an eventual slow-down as the cell approaches the source (Fig~\ref{fig:Fig3}A-C).

Realistic gradients are typically limited in spatial extent and often are not exponential, in which case $L$ and therefore $\tau_E$ \textcolor{black}{are} different in different regions. $L$ is long near the source in a linear gradient, for example, and decreases linearly with distance from the source. Simulations show that the cell initially climbs the gradient fast but later slows down as the gradient length scale $L$ increases and $\tau_E$ increases (Fig~\ref{fig:Fig3}D-F). On the contrary, for a static localized source in three dimensions, $L$ is short near the source but increases linearly with distance from it (Methods). Thus, $\tau_E$ decreases and the cell accelerates as it approaches the source (Fig~\ref{fig:Fig3}G-I). Comparing cells in various dynamical regimes (different values of $\tau_E$) across these different gradients suggests that a lower value of $\tau_E$ results in faster gradient ascent.

When entering a food gradient, it is natural to try to climb as fast as possible. However this strategy could create a problem: the longer runs implied by the positive feedback mechanism could propel the accelerating \textit{E. coli} beyond the nutrient source. This is the case in Fig.~\ref{fig:Fig3}E, where cells with the lowest $\tau_E$ (green) reach the source first but overshoot slightly; they settle, on average, at a further distance than those with intermediate $\tau_E$ (blue). \textcolor{black}{Thus there is a trade-off between transient gradient climbing and long-term aggregating, as previously observed~\cite{nick, yann, damon}. In nature, as chemotactic bacteria live in swarms, chasing and eating nutrient patches driven by flows and diffusions while new plumes of nutrients are constantly created by other organisms~\cite{marineMicrobes, blackburn}, the actual environments experienced by bacteria are far more complex. The trade-off we found here hints that in these random small fluctuating gradients~\cite{frequencyDependent,physicalLimits,vergassola,predictEnv,tindall} the bacteria should not aim for maximal drift speed but need to deal with this trade-off to avoid overshoot.} In general, natural environments will be complex, with a variety of different sources and gradients, implying different parameter domains will be optimal for \textit{E. coli} at different times. Such phenotypic diversity may well confer an advantage~\cite{nick,elowitz,ackermann,kussell,oudenaarden,tindall}.

\section*{Discussion}

Our results illustrate the surprisingly new capabilities that can emerge when living systems exploit the full nonlinearity \textcolor{black}{inherent within an otherwise simple and widely used} strategy. For the particular case of bacterial chemotaxis we showed that \textcolor{black}{cells} that swim fast, have long memory (adaptation time), or large signal amplification, are likely to \textcolor{black}{exhibit ``ratchet-like'' climbing behavior in a positive-feedback-dominated regime}, even in shallow gradients. \textcolor{black}{As we showed from simulations using a model that fits experimental data, this regime should be accessible to wild-type bacterial populations. Actually identifying these ``ratcheting'' cells from experimental trajectories would require observing them for a sufficient time ($T \gg t_M, t_D, t_E$) and in a sufficiently steep gradient over the distance traveled ($\Delta X = V_D T \sim 0.5 v_0 T$). Using parameter estimates from~\cite{yann,michael}, for $t_E < t_D \simeq t_M \sim 10~s$ we take $T \sim 200~s$, and for $v_0 = 20~\mu m/s$ we get $\Delta X = 2~mm$. To see how this compares with existing experimental setup with a quasi-linear gradient varying from $0$ to $1~mM$ over $10~mm$~\cite{adam}, we note that the black dots in Fig~\ref{fig:Fig1}A show that some cells located $1.5~mm$ away from $0$ concentration can operate in the positive-feedback-dominated regime. Thus, using the same setup as in~\cite{adam} these requirements would be satisfied near the bottom of the gradient if the source concentration was increased to $3~mM$.}

It is common to make simplifying assumptions to facilitate analysis, but we do not believe that ours are limiting. We showed with simulations that our results hold (\nameref{S1_Fig} for details) when we take into account: (i) different values of $r_0$ and $D_T/D_R$; (ii) the limited range of the receptor sensitivity~\cite{nick,tuReview} (\nameref{S1} Eq~(S10)); (iii) possible nonlinearities (\nameref{S1} Eq~(S4)) and asymmetries of adaptation rates~\cite{shimizu-tu-berg,asym-sourjik-endres}. A hallmark of \textit{E. coli} chemotaxis is that, in \textcolor{black}{the} absence of a gradient, run-and-tumble behavior adapts back to prestimulus statistics~\cite{bergCell,doyle}. These robust properties of integral feedback control~\cite{doyle} remain in place in our study because the transients originate from non-normal dynamics around the stable fixed point. The boost from positive feedback described here is independent from other mechanisms that can enhance drift up a gradient \textcolor{black}{such as imperfect adaptation in the response to some amino acids~\cite{vergassolaRace} and stochastic fluctuations in the adaptation mechanism~\cite{michael,tenwolde}. The latter has been shown to enhance chemotactic performance in shallow gradients by transiently pushing the system into a regime of slower direction changing provided it is running up the gradient. There are some similarities between the effect of signaling noise and the positive feedback mechanism presented here: both can affect drift speed by causing long-lasting asymmetries in the internal state when running up the gradient. In \nameref{S3_Fig} we show using simulations that signaling noise in the adaptation mechanism does not change our conclusion that the drift speed is maximal in the positive-feedback-dominated regime. Depending on the region of the $(\tau_E,\tau_D)$ parameter space, the signaling noise can either enhance the drift speed by less than $10 \%$ or reduce it by up to $30 \%$ .} 

The fact that non-normal dynamics might be exploited to boost runs in the correct direction parallels recent findings in neuroscience~\cite{neuron} that suggest neuronal networks use similar strategies to selectively amplify neural activity patterns of interest. Thus, non-normal dynamics could be a feature that is selected for in living dynamical systems. Although we used bacterial chemotaxis as an example, our results do not depend on the specific form of the functions $r(f)$ and $t_D(f)$\textcolor{black}{, provided} they are increasing. Therefore our findings should be applicable to a large class of biased random walk strategies exhibited by organisms when local directional information is unreliable. In essence, any stochastic navigation strategy requires a memory, $t_M$, to make temporal comparisons, a reorientation mechanism, $t_D$, to sample new directions, and external information, $t_E$, relayed to decision-making circuitry through motion and signal amplification. Our theoretical contribution showed the (surprisingly) diverse behavioral repertoire that is possible by having these work in concert. In retrospect, perhaps this should not be surprising given the diverse environments in which running-and-tumbling organisms can thrive.

\begin{table}[!h]
	\centering
	\caption{
		\textcolor{black}{		{\bf Symbol definitions.}}}
	\begin{tabular}{ | l | l | }
		\thickhline
		\multicolumn{2}{|l|}{\textcolor{black}{\bf Parameters}                                                       }\\ \thickhline
		\textcolor{black}{Name 		        }& \textcolor{black}{Definition                                                            }\\ \thickhline
		\textcolor{black}{$t_M$              }& \textcolor{black}{Memory, reciprocal to negative feedback                               }\\ \hline
		\textcolor{black}{$N$                }& \textcolor{black}{Receptor gain                                                         }\\ \hline
		\textcolor{black}{$H$                }& \textcolor{black}{Motor gain in $r(F) = 1/(1+\exp(-HF))$                                }\\ \hline
		\textcolor{black}{$F_0, f_0$         }& \textcolor{black}{Adapted internal state, $f_0=H F_0$                                   }\\ \hline
		\textcolor{black}{$r_0$              }& \textcolor{black}{Adapted probability to run $r(F_0)$                                   }\\ \hline
		\textcolor{black}{$\tau_{D0} $       }& \textcolor{black}{$t_D(F_0)/t_M$                                                        }\\ \hline
		\textcolor{black}{$v_0$              }& \textcolor{black}{Run speed                                                             }\\ \hline
		\textcolor{black}{$D_R$              }& \textcolor{black}{Rotational diffusion coefficient during runs                          }\\ \hline
		\textcolor{black}{$D_T$              }& \textcolor{black}{Rotational diffusion coefficient during tumbles                       }\\ \hline
		\textcolor{black}{$n$                }& \textcolor{black}{Spatial dimension                                                     }\\ \hline		
		\textcolor{black}{$K_a$              }& \textcolor{black}{Dissociation constant of receptor active state                        }\\ \hline
		\textcolor{black}{$K_i$              }& \textcolor{black}{Dissociation constant of receptor inactive state                      }\\ \thickhline
		\multicolumn{2}{|l|}{\textcolor{black}{\bf Independent Variables}                                            }\\ \thickhline
		\textcolor{black}{Name               }& \textcolor{black}{Definition                                                            }\\ \thickhline
		\textcolor{black}{${\bf X}, X, x$    }& \textcolor{black}{Position: vector, along gradient, $x = X/(v_0 t_M)$                   }\\ \hline
		\textcolor{black}{$t, \tau$          }& \textcolor{black}{Time, $\tau = t/t_M$                                                  }\\ \hline
		\textcolor{black}{$F, f$             }& \textcolor{black}{Internal state, $f=HF$                                                }\\ \hline
		\textcolor{black}{$s$                }& \textcolor{black}{Swimming direction $s= \hat{\bf{u}} \cdot \hat{\bf{X}} = \cos \theta$ }\\ \hline
		\textcolor{black}{$r$                }& \textcolor{black}{Probability to run $r(f) = 1/(1+\exp(-f))$                            }\\ \hline
		\textcolor{black}{$v$                }& \textcolor{black}{Normalized expected speed projected along gradient $v = r s$          }\\ \thickhline
		\multicolumn{2}{|l|}{\textcolor{black}{\bf Dependent Variables}                                              }\\ \thickhline
		\textcolor{black}{Name               }& \textcolor{black}{Definition                                                            }\\ \thickhline
		\textcolor{black}{$C({\bf X}, t)$    }& \textcolor{black}{Signal concentration                                                  }\\ \hline
		\textcolor{black}{$\phi({\bf X}, t)$ }& \textcolor{black}{Perceived signal $\ln ((1+C/K_i)/(1+C/K_a))$                          }\\ \hline
		\textcolor{black}{$L({\bf X})$       }& \textcolor{black}{Gradient length scale $1 /||\nabla \phi({\bf X})||$                   }\\ \hline
		\textcolor{black}{$\lambda_R(F)$     }& \textcolor{black}{Transition rate from run to tumble                                    }\\ \hline
		\textcolor{black}{$\lambda_T(F)$     }& \textcolor{black}{Transition rate from tumble to run                                    }\\ \hline
		\textcolor{black}{$t_S(F)$           }& \textcolor{black}{Run-tumble switching timescale $1/(\lambda_R(F) + \lambda_T(F))$      }\\ \hline
		\textcolor{black}{$t_E({\bf X})$     }& \textcolor{black}{Positive feedback timescale $L({\bf X})/(NHv_0)$                      }\\ \hline
		\textcolor{black}{$t_D(F)$           }& \textcolor{black}{Direction decorrelation timescale $1/((n-1)(r(F) D_R + (1-r(F) D_T))$ }\\ \hline
		\textcolor{black}{$\tau_E({\bf X})$  }& \textcolor{black}{Ratio between negative and positive feedbacks $t_E({\bf X})/t_M$      }\\ \hline
		\textcolor{black}{$\tau_D(F)$        }& \textcolor{black}{Ratio between keeping direction and memory $t_D(F)/t_M$               }\\ \hline
		\textcolor{black}{$P(x,f,s,\tau)$    }& \textcolor{black}{Probability distribution of the independent variables                 }\\ \hline
		\textcolor{black}{$p(f)$             }& \textcolor{black}{Marginal probability distribution of the internal state $f$           }\\ \thickhline
	\end{tabular}
	\label{tab:symbols}
\end{table}

\section*{Methods}

\subsection*{Agent-based simulation}

\subsubsection*{Chemotaxis pathway model}
A detailed description of the chemotaxis model and agent-based simulations is provided in \nameref{S1} (parameters in \nameref{Sparam}). Briefly, the agent-based simulations were performed using Euler's method to integrate a standard model of \textit{E. coli} chemotaxis~\cite{tuReview,shimizu-tu-berg,tuyuhai,yann,michael} in which the cell relays information from the external environment to the flagellar motor through a signaling cascade triggered by ligand-binding receptors. The receptors are described by a two-state model where the activity $a$ is determined by the free energy difference $F$ between the active and inactive states, which is determined by both the ligand concentration $C$ and the receptors' methylation level $m$. At each time step, the cell moves forward or stays in place according to its motility state (run or tumble), which also determines whether its direction changes with rotational diffusion coefficients $D_R$ or $D_T$. At the new position changes in ligand concentration $C$ cause changes in free energy $F$ and thus activity $a$, and the methylation state adapts to compensate that change to maintain a constant activity. The updated value of the free energy $F$ then determines the switching rates between the clockwise and counter-clockwise rotation of the flagellar motor state, which in turn determines the motility state of the cell according to rules and parameters in~\cite{michael}, completing one time step.

\subsubsection*{Noisy gene expression model}
In Fig~\ref{fig:Fig1}A we considered a wild type population in the scatter plot. To generate a population with realistic parameters, we used a recent model~\cite{adam} of phenotypic diversity in \textit{E. coli} chemotaxis that reproduces available experimental data on the laboratory strain RP437 climbing a linear gradient of methyl-aspartate. In this model individual cells have different abundances of the chemotaxis proteins (CheRBYZAW) and receptors (Tar, Tsr). These molecular abundances then determine the memory time $t_M$ and the adapted probability to run $r_0$~\cite{nick}. The run speed was different among cells and sampled from a Gaussian distribution to match experimental observations~\cite{adam}. Rotational diffusion coefficients were also distributed to reflect differences in cell length.

\subsection*{Derivation of Eqs~\eqref{eq:drdt}--\eqref{eq:FP}, the Fokker-Planck equation model in the fast switching limit}

\textcolor{black}{We define $P_R({\bf X}, \hat{\bf{u}}, F, t)$ and $P_T({\bf X}, \hat{\bf{u}}, F, t)$ as the probability distributions at time $t$ to be running or tumbling at position $\bf{X}$ in direction $\hat{\bf{u}}$ with internal variable $F$. As described, there is Poisson switching between runs and tumbles with rates $\lambda_R(F)$ and $\lambda_T(F)$, runs and tumbles follow rotational diffusion with $D_R$ and $D_T$, and motion is constant in runs and $0$ in tumbles. Thus we construct a two-state stochastic master equation model~\cite{stochastic}}

\begin{equation} \label{eq:master}
	\begin{aligned}
		\textcolor{black}{\partial_t P_R =}&\textcolor{black}{ - \partial _F \left( \dot{F}|_{run} P_R\right)  - \nabla \cdot \left( v_0 \hat{\bf{u}} P_R \right) +\nabla_{\hat{\bf{u}}}^2 \left( D_R P_R\right) - \lambda_R P_R +\lambda_T P_T}\\
		\textcolor{black}{\partial_t P_T =}&\textcolor{black}{ - \partial _F \left( \dot{F}|_{tumble} P_T \right) + \nabla_{\hat{\bf{u}}}^2 \left( D_T P_T\right) 
			+ \lambda_RP_R - \lambda_T  P_T,}
	\end{aligned}
\end{equation}
\textcolor{black}{where $\dot{F}|_{run,tumble}$ are defined in Eq~\eqref{eq:dFdtrt}.}

\textcolor{black}{Since the gradient varies in one direction only we focus on motion in the gradient direction and integrate the probability over all other directions. Thus $\nabla \cdot \hat{\bf{u}} = s \partial_X$ and $\nabla_{\hat{\bf{u}}}^2= \hat{L}_s$, the polar angle part of the rotational diffusion operator on the ($n-1$)-sphere. To derive the analytical form of $\hat{L}_s$ we note in $n$-dimensional space we can iteratively write down the Laplace-Beltrami operator~\cite{geometry} as}

\begin{equation} \label{eq:Laplace-Beltrami-general}
	\begin{aligned}
		\textcolor{black}{\nabla^2_{S^{n-1}} = \left(\sin\theta\right)^{2-n} \partial_\theta \left( \left(\sin\theta\right)^{n-2} \partial_\theta  \right) + \left(\sin\theta\right)^{-2}\nabla^2_{S^{n-2}},}
	\end{aligned}
\end{equation}
\textcolor{black}{where $0<\theta<\pi$ is the polar angle. In a one-dimensional gradient we define the gradient direction as the polar axis, thus $s= \hat{\bf{u}} \cdot \hat{\bf{X}} = \cos \theta$. We can write $\sin\theta = \sqrt{1-s^2}$ and $\partial_\theta=-\sqrt{1-s^2}\partial_s$. Then the polar angle part is}

\begin{equation} \label{eq:Laplace-Beltrami}
	\begin{aligned}
		\textcolor{black}{\hat{L}_s = \left(1-s^2\right)^{\frac{3-n}{2}} \partial_s \left( \left(1-s^2\right)^{\frac{n-1}{2}} \partial_s \right).}
	\end{aligned}
\end{equation}

\textcolor{black}{Using the definitions of the normalized internal state $f=HF$, of the timescale of switching between runs and tumbles $t_S=1/\left(\lambda_R+\lambda_T\right)$~\cite{stochastic}, and of the probability to run $r=\lambda_T / \left(\lambda_R+\lambda_T\right)$, we obtain }
\begin{equation} \label{eq:master-rescale}
	\begin{aligned}
		\partial_t P_R =&  - \partial _f \left( \left(-\frac{f-f_0}{t_M} + \frac{N H v_0}{L} s \right)P_R\right) +D_R \hat{L}_s P_R \\
		&- \frac{1-r}{t_S} P_R + \frac{r}{t_S} P_T - s\partial_X \left( v_0 P_R \right)\\
		\partial_t P_T =& - \partial _f \left(-\frac{f-f_0}{t_M}  P_T \right) + D_T \hat{L}_s P_T  \\
		&+ \frac{1-r}{t_S} P_R - \frac{r}{t_S}  P_T.
	\end{aligned}
\end{equation}

If we assume the switching terms with $t_S^{-1}$ in Eq~\eqref{eq:master-rescale} dominate, the probabilities to be running and tumbling equilibrate on a much faster timescale than the other ones. Therefore we can let $P=P_R+P_T$ and can approximate the actual probability to run as $ P_R/P \approx r$. Adding the two equations above yields the Fokker-Planck equation:
\begin{equation} \label{eq:FP-prescale}
	\begin{aligned}
		\partial_t P \approx& - \partial _f \left( \left(-\frac{f-f_0}{t_M} + \frac{r s}{L/NHv_0} \right)P\right) \\
		&+ \left(r D_R +\left(1-r\right) D_T\right) \hat{L}_s P - r s\partial_X \left( v_0 P \right).
	\end{aligned}
\end{equation}
This is equivalent to a system of Langevin equations. Considering $\mathrm{d}r/\mathrm{d}f = r(1-r)$ the internal variable dynamics (the first term on the right) gives Eq~\eqref{eq:drdt} which defines $t_E$. The angular dynamics (the second term on the right) defines $t_D$.

Using the time scale definitions in Eq~\eqref{eq:timescales} and non-dimensionalizing time $\tau = t/t_M$ and position $x=X/(v_0 t_M)$, we obtain the Fokker-Planck Eq~\eqref{eq:FP}.

\subsection*{Derivation of the drift speed $V_D$, Eq~\eqref{eq:VD}, \textcolor{black}{for a log-sensing organism moving in an exponential gradient}}

From the Fokker-Planck Eq~\eqref{eq:FP} we consider the steady state \textcolor{black}{so that $\partial_{\tau}=0$. For a log-sensing organism moving in an exponential gradient $\tau_E$ does not depend on $x$. We can therefore integrate over $x$ to get an equation for the marginal steady state distribution $\overline{P}(f,s)$ --- this removes the $\partial_x$ term.} Integrating over $s$ gives
\begin{equation} \label{eq:steady-state}
	\begin{aligned}
		0 = &- \partial_f \left( - \left(f-f_0\right)\int\overline{P}w(s)\mathrm{d}s  + \frac{r(f)}{\tau_E} \int s\overline{P}w(s)\mathrm{d}s \right),
	\end{aligned}
\end{equation}
where the bar indicates steady state. By the boundary conditions that $P\rightarrow0$ at $\pm\infty$, we must have
\begin{equation} \label{eq:p1p0balance}
	\begin{aligned}
		r(f)\int s\overline{P}w(s)\mathrm{d}s=\tau_E \left(f-f_0\right)\int\overline{P}w(s)\mathrm{d}s.
	\end{aligned}
\end{equation}
From the $-\partial_x\left(rsP\right)$ term of the Fokker-Planck Eq~\eqref{eq:FP}, the spatial flux is $r(f)s$ and the drift speed is its average over the distribution. Thus we get the drift speed as Eq~\eqref{eq:VD}
\begin{equation} \label{eq:VDderive}
	V_D = \overline{\langle rs\rangle} = \iint r(f) s \overline{P} w(s)\mathrm{d}s \mathrm{d}f = \tau_E\iint \left(f-f_0\right) \overline{P} w(s)\mathrm{d}s \mathrm{d}f=\tau_E \overline{\langle f-f_0\rangle}.
\end{equation}

\subsection*{Derivation of the analytical solution to the Fokker-Planck Eq~\eqref{eq:FP} by angular moment expansion \textcolor{black}{when $\tau_{D0} \ll 1$}}

\textcolor{black}{Here we use separation of variables and expand the solution to the Fokker-Planck Eq~\eqref{eq:FP} as a sum of eigenfunctions of the operator $\hat{L}_s$ on $s$. We then ignore high order terms assuming $\tau_{D0} \ll 1$ and derive an approximate analytical solution.}

\textcolor{black}{The eigenvalue problem of the angular operator $\hat{L}_s$, defined in Eq~\eqref{eq:Laplace-Beltrami}, is}

\begin{equation} \label{eq:Gegenbauer-ODE}
	\begin{aligned}
		\textcolor{black}{(1-s^2) y'' - (n-1) s y' = \lambda y.}
	\end{aligned}
\end{equation}
\textcolor{black}{We identify this as the Gegenbauer differential equation~\cite{Gegenbauer}, with eigenfunctions the Gegenbauer polynomials $C_k^{(n/2-1)}(s)$ and the corresponding eigenvalues $\lambda_k^{(n/2-1)} = -k(k+n-2)$. When $n=3$ they are Legendre polynomials with eigenvalues $\lambda_k^{(1/2)} = -k\left(k+1\right)$. The first few Gegenbauer polynomials are}

\begin{equation} \label{eq:Gegenbauer-poly}
	\begin{aligned}
		\textcolor{black}{C_0^{(n/2-1)}(s) }&\textcolor{black}{= 1}\\
		\textcolor{black}{C_1^{(n/2-1)}(s) }&\textcolor{black}{= \left(n-2\right)s}\\
		\textcolor{black}{C_2^{(n/2-1)}(s) }&\textcolor{black}{= \frac{n-2}{2}\left(ns^2 - 1\right).}
	\end{aligned}
\end{equation}
\textcolor{black}{They are orthogonal in the sense that}
\begin{equation} \label{eq:GegenOrtho}
	\begin{aligned}
		\textcolor{black}{\int_{-1}^1 C_k^{(n/2-1)}(s) C_l^{(n/2-1)}(s) \left(1-s^2\right)^{\frac{n-3}{2}} \mathrm{d}s = N_k^{(n/2-1)},}
	\end{aligned}
\end{equation}
\textcolor{black}{where the normalization constants are $N_k^{(n/2-1)}=\frac{\pi 2^{4-n}(k+n-3)!}{k!(2k+n-2)\left(\Gamma(n/2-1)\right)^2}$. When $n=3$ they are $N_k^{(1/2)}=\frac{2}{2k+1}$, those of Legendre polynomials.}

\textcolor{black}{The weight in the integration above is consistent with the geometry on an $(n-1)$-sphere $S^{n-1}$, whose the volume element are iteratively defined~\cite{geometry} as}

\begin{equation} \label{eq:vol-el}
	\begin{aligned}
		\textcolor{black}{\mathrm{d}_{S^{n-1}}\omega = \left(\sin\theta\right)^{n-2}\mathrm{d}\theta\mathrm{d}_{S^{n-2}}\omega.}
	\end{aligned}
\end{equation}
\textcolor{black}{After a change of variable $s=\cos\theta$ and integrating over all remaining dimensions, we see that any integration of $s$ should carry a weight}
\begin{equation} \label{eq:weight}
	\begin{aligned}
		\textcolor{black}{w(s)\mathrm{d}s = \left(1-s^2\right)^{\frac{n-3}{2}} \mathrm{d}s.}
	\end{aligned}
\end{equation}

\textcolor{black}{From orthogonality and completeness, we write any function of $s$, in particular the probability distribution $P$, as a series of Gegenbauer polynomials. When $n=3$ this is the Fourier-Legendre Series.}
\begin{equation} \label{eq:FLSeries}
	\begin{aligned}
		\textcolor{black}{P(x,f,s,\tau)~}&\textcolor{black}{ = \sum_{k=0}^{\infty} p_k(x,f,\tau)\frac{C_k^{(n/2-1)}(s)}{\sqrt{N_0^{(n/2-1)}N_k^{(n/2-1)}}} }\\
		&\textcolor{black}{ = \frac{1}{N_0^{(n/2-1)}}\left( p_0 + p_1 \sqrt{n} s + p_2 \sqrt{\frac{n+2}{n-1}}\frac{n s^2-1}{2} + \cdots\right),}\\
		\textcolor{black}{p_k(x,f,\tau)~}&\textcolor{black}{= \int_{-1}^{1} \sqrt{\frac{N_0^{(n/2-1)}}{N_k^{(n/2-1)}}} C_k^{(n/2-1)}(s)  P(x,f,s,\tau)\left(1-s^2\right)^{\frac{n-3}{2}} \mathrm{d}s,}
	\end{aligned}
\end{equation}
\textcolor{black}{where we normalize the definitions to ensure $p_0 = \int_{-1}^{1} P\left(1-s^2\right)^{\frac{n-3}{2}} \mathrm{d}s$ is the same as the marginal distribution. When $n=3$, the above is}

\begin{equation} \label{eq:FLSeries3}
	\begin{aligned}
		\textcolor{black}{P(x,f,s,\tau)~}&\textcolor{black}{ = \sum_{k=0}^{\infty} p_k(x,f,\tau)\frac{\sqrt{2k+1}}{2}C_k^{(1/2)}(s) }\\
		&\textcolor{black}{ = \frac{1}{2}\left( p_0 + p_1 \sqrt{3} s + p_2 \sqrt{\frac{5}{2}}\frac{3 s^2-1}{2} + \cdots\right),}\\
		\textcolor{black}{p_k(x,f,\tau)~}&\textcolor{black}{= \int_{-1}^{1} \sqrt{2k+1} C_k^{(1/2)}(s)  P(x,f,s,\tau) \mathrm{d}s.}
	\end{aligned}
\end{equation}
	
\textcolor{black}{From now on we denote the marginal distribution $p(f)=p_0(f)$. Also, from this definition $p_1 =\sqrt{n} \int_{-1}^{1} s P\left(1-s^2\right)^{\frac{n-3}{2}} \mathrm{d}s$.}

\textcolor{black}{Substitute the expansion Eq~\eqref{eq:FLSeries} into the Fokker-Planck Eq~\eqref{eq:FP} and use the orthogonality Eq~\eqref{eq:GegenOrtho}, we obtain}
\begin{equation} \label{eq:pkPDEs}
	\begin{aligned}
		\partial_\tau p_k =& -\partial_f \left(-\left(f-f_0\right) p_k+ \frac{r(f)}{\tau_E} \hat{s}_{kl} p_l\right) + \frac{ \lambda_k^{(n/2-1)}}{(n-1)\tau_D(f)} p_k - \partial_x\left( \hat{s}_{kl} p_l\right),
	\end{aligned}
\end{equation}
where $\hat{s}_{kl}= \sqrt{\frac{k\left(k+n-3\right)}{\left(2k+n-4\right)\left(2k+n-2\right)}}\delta_{k-1,l}+ \sqrt{\frac{\left(k+1\right)\left(k+n-2\right)}{\left(2k+n-2\right)\left(2k+n\right)}} \delta_{k+1,l}$ (summation over $l$ implied) is an operator relating neighboring orders. It comes from the positive feedback term. When $n=3$ it is $\hat{s}_{kl}= \frac{k}{\sqrt{4k^2-1}} \delta_{k-1,l}+\frac{k+1}{\sqrt{4\left(k+1\right)^2-1}} \delta_{k+1,l}$. \textcolor{black}{The first few equations are}

\begin{equation} \label{eq:firstFewPk}
	\begin{aligned}
		\textcolor{black}{\partial_{\tau} p = }&\textcolor{black}{- \partial_f \left( - \left(f-f_0\right) p + \frac{r(f)}{\tau_E} \frac{1}{\sqrt{n}} p_1 \right) -r(f) \partial_{x} \frac{1}{\sqrt{n}} p_1}\\
		\textcolor{black}{\partial_{\tau}  p_1 = }&\textcolor{black}{ - \partial_f \left(  - \left(f-f_0\right) p_1 + \frac{r(f)}{\tau_E} \left(\frac{1}{\sqrt{n}}p+\sqrt{\frac{2(n-1)}{n\left(n+2\right)}}p_2\right) \right)}\\
		&\textcolor{black}{  - \frac{1}{\tau_D(f)}p_1-r(f) \partial_{x}\left(\frac{1}{\sqrt{n}}p+\sqrt{\frac{2(n-1)}{n\left(n+2\right)}}p_2\right)}\\
		\textcolor{black}{\partial_{\tau}  p_2 = }&\textcolor{black}{ - \partial_f \left(  - \left(f-f_0\right) p_2 + \frac{r(f)}{\tau_E} \left(\sqrt{\frac{2(n-1)}{n\left(n+2\right)}}p_1+\sqrt{\frac{3n}{\left(n+2\right)\left(n+4\right)}}p_3\right) \right)}\\
		&\textcolor{black}{  - \frac{2n}{(n-1)\tau_D(f)}p_2-r(f) \partial_{x}\left(\sqrt{\frac{2(n-1)}{n\left(n+2\right)}}p_1+\sqrt{\frac{3n}{\left(n+2\right)\left(n+4\right)}}p_3\right).}
\end{aligned}
\end{equation}

\textcolor{black}{In the definition of $\hat{s}_{kl}$, when $k \gg 1$ the non-zero entries approach a constant $1/2$. This means for large $k$ the coefficients $p_k$ in Eq~\eqref{eq:pkPDEs} evolve similarly except that higher orders decay with faster rates $k\left(k+n-2\right) / (n-1)\tau_D$. Therefore when $\tau_{D0} \ll 1$ we can neglect the 2nd and higher orders, which closes the infinite series of moment equations and leaves two equations concerning the zeroth and first marginal moments in $s$, $p(x,f,\tau)$ and $p_1(x,f,\tau)$ respectively.} At steady state the approximation gives the analytical solution
\begin{equation} \label{eq:dpf}
	\begin{aligned}
		\overline{p}(f)=&  \frac{1}{W}\frac{\frac{r(f)}{\tau_E}}{\frac{1}{n} \left(\frac{r(f)}{\tau_E}\right)^2 - \left(f-f_0\right)^2} \exp\left( - \int^f \frac{ \frac{f_1-f_0}{\tau_D(f_1)} }{\frac{1}{n} \left(\frac{r(f_1)}{\tau_E}\right)^2 - \left(f_1-f_0\right)^2} \mathrm{d} f_1 \right),
	\end{aligned}
\end{equation}
where $W$ is a normalization constant. \textcolor{black}{Eq~\eqref{eq:dpf} is the same as Eq~\eqref{eq:dpf_main} in the main text.}

\textcolor{black}{We can interpret the steady state distribution as a potential solution $\overline{p}(f) \propto \exp -V(f)$ where $V(f)$ is the ``potential''. In this case the equivalent ``force'' in internal state is}

\begin{equation} \label{eq:fForce}
	\begin{aligned}
		\textcolor{black}{F(f) =}&\textcolor{black}{ -V'(f) = \frac{\mathrm{d}\ln\overline{p}(f)}{\mathrm{d}f}}\\
		\textcolor{black}{=}&\textcolor{black}{ \frac{\mathrm{d} }{\mathrm{d}f} \ln\frac{\frac{r(f)}{\tau_E}}{\frac{1}{n} \left(\frac{r(f)}{\tau_E}\right)^2 - \left(f-f_0\right)^2} - \frac{ \frac{f-f_0}{\tau_D(f)} }{\frac{1}{n} \left(\frac{r(f)}{\tau_E}\right)^2 - \left(f-f_0\right)^2}.}
	\end{aligned}
\end{equation}
\textcolor{black}{Since $\tau_{D0} \ll 1$ the second term dominates, making the ``force'' a spring-like system, with spring constant}
\begin{equation} \label{eq:spring}
	\textcolor{black}{k(f) = \frac{ \frac{1}{\tau_D(f)} }{\frac{1}{n} \left(\frac{r(f)}{\tau_E}\right)^2 - \left(f-f_0\right)^2}.}
\end{equation}
\textcolor{black}{Three observations can be made from this spring constant in intuitively understanding the steady state distribution $\overline{p}(f)$. (i) $k(f) \rightarrow \infty$, i.e. the ``spring'' becomes infinitely ``stiff'', when the denominator approaches 0. Therefore, the bounds of the distribution $\overline{p}(f)$ are proportional to $1/\tau_E$, the ratio between the positive and negative feedbacks (Eq~\eqref{eq:timescales}). Intuitively, a stronger positive feedback (smaller $\tau_E$) drives the internal state $f$ further away from $f_0$, so the spring constant $k(f)$ is smaller and the distribution $\overline{p}(f)$ is wider. (ii) A slower change in direction (smaller $\tau_D$) leads to a larger spring constant $k(f) \propto 1/\tau_D(f)$, and thus the distribution $\overline{p}(f)$ is more concentrated near the ``origin'' $f_0$. Intuitively, a shorter direction correlation time $\tau_D$ inhibits coherent motion in a single direction, which is required by the positive feedback to consistently drive the internal state $f$ away. Thus the distribution $\overline{p}(f)$ is more concentrated. (iii) Asymmetries are created by the functional dependencies of $r(f)$ and $\tau_D(f)$, both increasing in $f$ --- a ``weaker spring'' for higher values of $f$ shifts the distribution $\overline{p}(f)$ there. Intuitively, more positive feedback $\propto r(f)$ and more coherent motion $\propto \tau_D(f)$ in the positive direction asymmetrically drives the internal state towards higher values. These 3 observations can all be found in Fig~\ref{fig:Fig1}C.}

\subsection*{Derivation of the distribution $\overline{p}(f)$ and drift speed $V_D$ when the negative feedback dominates.}

\textcolor{black}{We expand the steady state solution Eq~\eqref{eq:dpf} in orders of $\frac{1}{\tau_E} \ll 1$ and $\tau_{D0} \ll 1$ and obtain a near-Gaussian approximation, from which we integrate using Eq~\eqref{eq:VD} to obtain MFT results.}

\textcolor{black}{First, we write the steady state distribution Eq~\eqref{eq:dpf} as}

\begin{equation} \label{eq:pfAB}
	\begin{aligned}
		\textcolor{black}{\overline{p}(f) = \frac{1}{W} B(f) \exp \left( - \int^f A(f_1)\mathrm{d}f_1\right).}
	\end{aligned}
\end{equation}
\textcolor{black}{From the Taylor expansion of the integrand in the exponent}
\begin{equation} \label{eq:ATaylor}
	\begin{aligned}
		\textcolor{black}{A(f) =}&\textcolor{black}{ \frac{\frac{f-f_0}{\tau_D(f)} }{\frac{1}{n} \left(\frac{r(f)}{\tau_E}\right)^2 - \left(f-f_0\right)^2} }\\
		\textcolor{black}{=}&\textcolor{black}{ \frac{n\tau_E^2}{r_0^2\tau_{D0}}\left(f-f_0\right) + \left(1+O(\frac{1}{\tau_E^2})\right)  \Sigma_{m=1}^\infty \frac{n^{m+1}\tau_E^{2m+2}}{r_0^{2m+2}\tau_{D0}}\left(f-f_0\right)^{2m+1} }\\
		&\textcolor{black}{ -\left(1+O(\frac{1}{\tau_E^2})\right)  \Sigma_{m=1}^\infty \frac{n^{m}\tau_E^{2m}r_0'}{r_0^{2m+1}\tau_{D0}}\left(2m+\frac{r_0\tau_{D0}'}{r_0'\tau_{D0}}\right)\left(f-f_0\right)^{2m},}
	\end{aligned}
\end{equation}
\textcolor{black}{where $' = \mathrm{d}/\mathrm{d}f$, we see that if we define}

\begin{equation} \label{eq:variance}
	\textcolor{black}{\sigma^2 = \frac{r_0^2\tau_{D0}}{n\tau_E^2},}
\end{equation}
\textcolor{black}{the first term in $A(f)$ will give $-\int A(f)\mathrm{d}f = -\frac{\left(f-f_0\right)^2}{2\sigma^2}+...$. If we can show that the rest of the terms are small when $\frac{1}{\tau_E} < 1$ and $\tau_{D0} < 1$, we can write $\overline{p}(f)$ as a small deviation from a Gaussian.}

\textcolor{black}{Indeed, if we consider the integration range $|f-f_0| \sim \sigma \sim O(\sqrt{\tau_{D0}}/\tau_E)$, we can write}

\begin{equation} \label{eq:AExpand}
	\begin{aligned}
		\textcolor{black}{-\int^f_{f_0} A(f_1) \mathrm{d}f_1 = }&\textcolor{black}{ - \frac{\left(f-f_0\right)^2}{2 \sigma^2} - \Sigma_{m=1}^\infty\frac{\tau_{D0}^m}{2m+2}\frac{\left(f-f_0\right)^{2m+2}}{\sigma^{2m+2}} }\\
		&\textcolor{black}{ + \Sigma_{m=1}^\infty \frac{r_0'}{r_0} \frac{\tau_{D0}^{m-1}}{2m+1}\left(2m+\frac{r_0\tau_{D0}'}{r_0'\tau_{D0}}\right) \frac{\left(f-f_0\right)^{2m+1}}{\sigma^{2m}} + O(\frac{1}{\tau_E^2})}\\
		\textcolor{black}{= }&\textcolor{black}{ - \frac{\left(f-f_0\right)^2}{2 \sigma^2} + \frac{r_0'}{r_0} \frac{1}{3}\left(2+\frac{r_0\tau_{D0}'}{r_0'\tau_{D0}}\right) \frac{\left(f-f_0\right)^3}{\sigma^2} -\frac{\tau_{D0}}{4}\frac{\left(f-f_0\right)^4}{\sigma^4} }\\
		&\textcolor{black}{ + \frac{r_0'}{r_0} \frac{\tau_{D0}}{5}\left(4+\frac{r_0\tau_{D0}'}{r_0'\tau_{D0}}\right) \frac{\left(f-f_0\right)^5}{\sigma^4} + O(\frac{1}{\tau_E^2}) + O(\tau_{D0}^2),}
	\end{aligned}
\end{equation}
\textcolor{black}{Similarly, the prefactor is}
\begin{equation} \label{eq:BExpand}
	\begin{aligned}
		\textcolor{black}{B(f) =}&\textcolor{black}{ \frac{n\tau_E}{r_0} \bigg(1 - \frac{r_0'}{r_0} \left(f-f_0\right) + \tau_{D0}\frac{\left(f-f_0\right)^2}{\sigma^2} - 3\tau_{D0} \frac{r_0'}{r_0} \frac{\left(f-f_0\right)^3}{\sigma^2}}\\
		&\textcolor{black}{ + O(\frac{1}{\tau_E^2}) + O(\tau_{D0}^2)\bigg).}
	\end{aligned}
\end{equation}

\textcolor{black}{Substitute Eqs~\eqref{eq:AExpand}-\eqref{eq:BExpand} back into Eq~\eqref{eq:pfAB} and taking care of the orders of all cross terms, we obtain}

\begin{equation} \label{eq:pexp}
	\begin{aligned}
		\textcolor{black}{\overline{p}(f)}&\textcolor{black}{ = \frac{1}{Z} \frac{ \mathrm{e}^{-\frac{\left(f-f_0\right)^2}{2\sigma^2}}}{\sqrt{2\pi\sigma^2}}  \bigg( 1- \frac{r_0'}{r_0} \left(f-f_0\right) + \tau_{D0}\frac{\left(f-f_0\right)^2}{\sigma^2} }\\
		&\textcolor{black}{ + \frac{r_0'}{3r_0}\left(2- 9\tau_{D0}+\frac{r_0\tau_{D0}'}{r_0'\tau_{D0}} \right) \frac{\left(f-f_0\right)^3}{\sigma^2} -\frac{\tau_{D0}}{4}\frac{\left(f-f_0\right)^4}{\sigma^4} }\\
		&\textcolor{black}{ + \frac{r_0'}{60r_0}\left(103+32\frac{r_0\tau_{D0}'}{r_0'\tau_{D0}} \right) \tau_{D0} \frac{\left(f-f_0\right)^5}{\sigma^4} }\\
		&\textcolor{black}{ - \frac{r_0'}{12r_0}\left(2+\frac{r_0\tau_{D0}'}{r_0'\tau_{D0}} \right) \tau_{D0} \frac{\left(f-f_0\right)^7}{\sigma^6} + O(\frac{1}{\tau_E^3}) + O(\tau_{D0}^{\frac{5}{2}}) \bigg).}
	\end{aligned}
\end{equation}
\textcolor{black}{with normalization constant $Z$.}

\textcolor{black}{We notice from Eq~\eqref{eq:pfAB} that the range of distribution is bounded by $f_L$ and $f_U$, defined by}

\begin{equation} \label{eq:fLU}
	\begin{aligned}
		\textcolor{black}{f_L - f_0 = }&\textcolor{black}{ - \frac{1}{\sqrt{n}}\frac{r(f_L)}{\tau_E},}\\
		\textcolor{black}{f_U - f_0 = }&\textcolor{black}{ \frac{1}{\sqrt{n}}\frac{r(f_U)}{\tau_E}.}
\end{aligned}
\end{equation}
\textcolor{black}{Since $\sigma = \frac{r_0 \sqrt{\tau_{D0}}}{\sqrt{n}\tau_E} \ll \frac{r_0}{\sqrt{n}\tau_E}$, we see that the integration range is much larger than the standard deviation of the Gaussian factor, and thus can be considered from $-\infty$ to $\infty$. Therefore we get the normalization constant}
\begin{equation} \label{eq:pnorm}
	\textcolor{black}{Z = 1 + \frac{\tau_{D0}}{4}  + O(\frac{1}{\tau_E^2}) + O(\tau_{D0}^2).}
\end{equation}

\textcolor{black}{Substitute Eq~\eqref{eq:pexp} and Eq~\eqref{eq:pnorm} into Eq~\eqref{eq:VD} and carry out the integrals}

\begin{equation} \label{eq:Vdexp0}
	\textcolor{black}{V_D = \frac{r_0\tau_{D0}}{n\tau_E} \frac{\left(1-\frac{3}{4}\tau_{D0}\right)\left(r_0'+r_0\frac{\tau_{D0}'}{\tau_{D0}}\right)+O(\frac{1}{\tau_E}) + O(\tau_{D0}^{\frac{3}{2}})}{1 + \frac{\tau_{D0}}{4}  + O(\frac{1}{\tau_E^2}) + O(\tau_{D0}^2)}.}
\end{equation}

\textcolor{black}{Finally, noticing that by the definition of $\tau_D$ in Eq~\eqref{eq:timescales}}
\begin{equation} \label{eq:tauD}
	\textcolor{black}{\tau_D(f) = \tau_{D0} \frac{r_0D_R+\left(1-r_0\right)D_T}{r(f)D_R+\left(1-r(f)\right)D_T},}
\end{equation}
\textcolor{black}{we can get}
\begin{equation} \label{eq:tauD'}
	\textcolor{black}{\tau_{D0}' = \tau_{D0} \frac{D_T-D_R}{r_0D_R+\left(1-r_0\right)D_T} r_0'.}
\end{equation}
\textcolor{black}{Therefore}
\begin{equation} \label{eq:r0'}
	\begin{aligned}
		\textcolor{black}{r_0'+r_0\frac{\tau_{D0}'}{\tau_{D0}} }&\textcolor{black}{ = \frac{\tau_{D0}'}{\tau_{D0}}\left(\frac{r_0D_R+\left(1-r_0\right)D_T}{D_T-D_R}+r_0\right) }\\
		&\textcolor{black}{ = \frac{\tau_{D0}'}{\tau_{D0}}\frac{D_T}{D_T-D_R}.}
\end{aligned}
\end{equation}
\textcolor{black}{Taking $D_T \gg D_R$, we put this back into Eq~\eqref{eq:Vdexp0} and get}
\begin{equation} \label{eq:VDMFT}
	\begin{aligned}
		\textcolor{black}{V_D }&\textcolor{black}{= \frac{r_0\tau_{D0}'}{n\tau_E} \frac{1-\frac{3}{4}\tau_{D0}+O(\frac{1}{\tau_E}) + O(\tau_{D0}^{\frac{3}{2}})}{1 + \frac{\tau_{D0}}{4}  + O(\frac{1}{\tau_E^2}) + O(\tau_{D0}^2)} }\\
		&\textcolor{black}{= \frac{r_0\tau_{D0}'}{n\tau_E\left(1+\tau_{D0}\right)} \left(1 + O(\frac{1}{\tau_E}) + O(\tau_{D0}^{\frac{3}{2}})\right).}
\end{aligned}
\end{equation}

\textcolor{black}{When converted back to real units ($t$ instead of $\tau = t/t_M$), the highest-order term is identical, except for notations, to Eq~(3) in Dufour \textit{et al.}~\cite{yann} obtained from a different approach. It can also be reduced to Eq~(12) in Si \textit{et al.}~\cite{tuyuhai} by assuming a high running probability and a long memory. It agrees with Eq~(6.24) in Erban \& Othmer~\cite{erban-othmer2004} and Eq~(16) in Franz \textit{et al.}~\cite{franz} with appropriate inclusion of rotational diffusion.}

\textcolor{black}{In Eq~\eqref{eq:Vdexp0} we expanded the distribution as a near-Gaussian around $f_0$. From Eq~\eqref{eq:VD} we see the mean internal state $f_m=\overline{\langle f\rangle}$ has a slight shift, so it's more accurate to expand around $f_m$. From Eq~\eqref{eq:Vdexp0} and Eq~\eqref{eq:VD} in the main text, we see $\overline{\langle f-f_0\rangle} \sim O(1/\tau_E^2) $. Thus considering the shift in $f_m$ the resulting $V_D$ has the same form compared to Eq~\eqref{eq:VDMFT}:}
\begin{equation} \label{eq:VDMFT_m}
	\textcolor{black}{V_D = \frac{r_m\tau_{Dm}'}{n\tau_E\left(1+\tau_{Dm}\right)} \left(1 + O(\frac{1}{\tau_E}) + O(\tau_{Dm}^{\frac{3}{2}})\right).}
\end{equation}

\subsection*{Bounds of the distribution $p(f)$}

The first term in Eq~\eqref{eq:FP} says the flux in $f$-space is non-negative provided $-\left(f-f_0\right) + rs/\tau_E > 0$, or, noting $s\le1$
\begin{equation} \label{eq:fupper}
	f \le f_0 + r(f) s/\tau_E \le f_0 + r(f) /\tau_E.
\end{equation}
Thus the upper bound $f_U$ of the distribution $p(f)$ is achieved at equality. Similarly, the lower bound $f_L$ is achieved when we take equal signs of
\begin{equation} \label{eq:flower}
	f \ge f_0 + r(f) s/\tau_E \ge f_0 - r(f) /\tau_E,
\end{equation}
noting $s\ge1$.

When $\tau_E$ becomes small we note $f_{U,L}$ deviates far away from $f_0$ as $1/\tau_E\rightarrow \infty$. Using the definition $r = 1/\left(1+\exp\left(-f\right)\right)$, we write
\begin{equation} \label{eq:fBounds}
	\begin{aligned}
		f_{L,U} = \mp \frac{1}{\tau_E\left(1+\exp\left(-f_{L,U}\right)\right)}.
	\end{aligned}
\end{equation}
The plus sign gives $\exp\left(-f_U)\right) \ll 1$ and $f_U \approx 1/\tau_E$. The minus sign gives $\exp\left(-f_L)\right) \gg 1$ and $f_L = - \exp\left(f_L\right)/\tau_E$. Taking logarithm, the latter gives $f_L = \ln \left(\left| f_L \right| \tau_E \right) \approx \ln \tau_E$.

\subsection*{\textcolor{black}{Derivation of Langevin Eq~\eqref{eq:Langevin}}}

\textcolor{black}{To derive Langevin equations from the Fokker-Planck equation we need to consider the geometric weight factor $w(s)$ in Eq~\eqref{eq:weight} for anglular integration. In deriving $s$-dynamics, we start with the angular part of the Fokker-Planck Eq~\eqref{eq:FP}}
\begin{equation} \label{eq:Pangular}
	\textcolor{black}{\partial_{\tau} P = \frac{\hat{L}_s P}{(n-1)\tau_D(f)} + \ldots.}
\end{equation}
\textcolor{black}{Multiplying an arbitrary function $A(s)$ and integrating over all dimensions, we obtain}
\begin{equation} \label{eq:As1}
	\begin{aligned}
		&\textcolor{black}{ \int\mathrm{d}x\int\mathrm{d}f \int_{-1}^1w(s)\mathrm{d}sA(s)\partial_{\tau} P}\\
		\textcolor{black}{= }&\textcolor{black}{ \int\mathrm{d}x\int\mathrm{d}f \int_{-1}^1w(s)\mathrm{d}sA(s) \frac{\left(1-s^2\right)^{\frac{3-n}{2}} \partial_s \left( \left(1-s^2\right)^{\frac{n-1}{2}} \partial_s P\right)}{(n-1)\tau_D(f)} }\\
		\textcolor{black}{= }&\textcolor{black}{ - \int\mathrm{d}x\int\mathrm{d}f \int_{-1}^1w(s)\mathrm{d}s \frac{sP}{\tau_D(f)} \partial_sA(s) }\\
		&\textcolor{black}{ + \int\mathrm{d}x\int\mathrm{d}f \int_{-1}^1w(s)\mathrm{d}s \frac{ \left(1-s^2\right)P}{(n-1)\tau_D(f)} \partial_s^2A(s).}
	\end{aligned}
\end{equation}

\textcolor{black}{To apply the standard result of equivalence between Fokker-Planck equations and Langevin equations, we need to change the measure in $s$-space to unity. This prompts the definition $Q(s,t)=w(s)\iint P(y,f,s,t) \mathrm{d}x \mathrm{d}f$ so that the above becomes}
\begin{equation} \label{eq:As2}
	\begin{aligned}
		\textcolor{black}{\int_{-1}^1\mathrm{d}sA(s)\partial_{\tau} Q = }&\textcolor{black}{- \int_{-1}^1\mathrm{d}s \frac{sQ}{\tau_D(f)} \partial_sA(s) + \int_{-1}^1\mathrm{d}s \frac{ \left(1-s^2\right)Q}{(n-1)\tau_D(f)} \partial_s^2A(s)}\\
		\textcolor{black}{= }&\textcolor{black}{ \int_{-1}^1\mathrm{d}s A(s) \partial_s\frac{sQ}{\tau_D(f)} + \int_{-1}^1\mathrm{d}s A(s)\partial_s^2 \frac{ \left(1-s^2\right)Q}{(n-1)\tau_D(f)},}
	\end{aligned}
\end{equation}
\textcolor{black}{where we integrated by parts and discarded boundary terms. Since $A(s)$ is an arbitrary function, we can write down the Fokker-Planck equation}
\begin{equation} \label{eq:QFPE}
	\textcolor{black}{\partial_{\tau} Q = \partial_s\frac{s Q}{\tau_D(f)} + \partial_s^2 \frac{ \left(1-s^2\right)Q}{(n-1)\tau_D(f)},}
\end{equation}
\textcolor{black}{which is equivalent~\cite{stochastic} to the Langevin equation}
\begin{equation} \label{eq:s-dyn}
	\textcolor{black}{\frac{\mathrm{d} s}{\mathrm{d}\tau} = - \frac{s}{\tau_D(r)} + \sqrt{\frac{2\left(1-s^2\right)}{(n-1)\tau_D(r)}} \eta(\tau).}
\end{equation}
\textcolor{black}{where $\eta(\tau)$ denotes the Gaussian white noise with $\langle\eta(\tau_1)\eta(\tau_2)\rangle=\delta(\tau_1-\tau_2)$.}

\textcolor{black}{The other two variables follow standard results~\cite{stochastic} from the Fokker-Planck Eq~\eqref{eq:FP} in the main text}
\begin{equation} \label{eq:f-dyn}
	\begin{aligned}
		\textcolor{black}{\frac{\mathrm{d} f }{\mathrm{d}\tau} }&\textcolor{black}{= - \left(f-f_0\right) + \frac{r(f) s}{\tau_E},}\\
		\textcolor{black}{\frac{\mathrm{d} x }{\mathrm{d}\tau} }&\textcolor{black}{=  r(f) s.}
\end{aligned}
\end{equation}

\textcolor{black}{Now we change variables according to the definitions $r(f) = 1/\left(1+\exp(-f)\right)$ and $v = rs$, and derive from the above dynamics in Eq~\eqref{eq:s-dyn} and Eq~\eqref{eq:f-dyn} to get the Langevin Eq~\eqref{eq:Langevin}.}

\subsection*{Linear response near the fixed point of the Langevin system}

Near the fixed point $(r_0,0)$, the eigenvectors and eigenvalues of the linearized Langevin Eq~\eqref{eq:Langevin} are: 
\begin{equation} \label{eq:LangevinEigen}
	\begin{aligned}
		\begin{bmatrix}
			1 \\
			0
		\end{bmatrix}
		&\text{~~~for eigenvalue~~~} -1;\\
		\begin{bmatrix}
			\frac{\left(1-r_0\right)r_0}{\tau_E}\frac{\tau_{D0}}{\tau_{D0}-1} \\
			1
		\end{bmatrix}
		&\text{~~~for eigenvalue~~~} -\frac{1}{\tau_{D0}}.
	\end{aligned}
\end{equation}
When $\tau_E$ is large, $\frac{\left(1-r_0\right)r_0}{\tau_E}\frac{\tau_{D0}}{\tau_{D0}-1} \ll 1$ and the eigenvectors are almost orthogonal. When $\tau_E$ is small, $\frac{\left(1-r_0\right)r_0}{\tau_E}\frac{\tau_{D0}}{\tau_{D0}-1} \gg 1$ and the eigenvectors are not orthogonal.

\subsection*{Numerical methods}

In Fig~\ref{fig:Fig1}A heat map the drift speed $V_D$ was calculated by fitting the linear part of the mean trajectory. In Fig~\ref{fig:Fig1}B the first $50 ~ s$ were removed to avoid the start up transient. In Fig~\ref{fig:Fig1}C, the steady state $\overline{p}(f)$ from agent-based simulations was calculated from the histogram of all the internal values of the $10^4$ simulated cells between $\tau = 10$ and $\tau = 20$, sampled at regular steps of $\tau = 0.01$. Numerical solutions of the Fokker-Planck Eq~\eqref{eq:FP} were obtained by expanding the distribution in angles, as in Eq~\eqref{eq:pkPDEs}, and keeping the first 10 orders. The steady state $\overline{p}(f)$ was found by solving an initial value problem using the NDSolve function in Mathematica, with $10^4$ spatial points and integration time up to $\tau = 10$. Further orders, finer grid, and longer integration times were checked to ensure solution accuracy. In Fig~\ref{fig:Fig1}D, $V_D$ from agent-based and Fokker-Planck were calculated by plugging into Eq~\eqref{eq:VD} $\overline{p}(f)$ obtained from those methods in C. MFT was calculated by combining Eq~\eqref{eq:VDMFT} with Eq~\eqref{eq:VD} to find both $f_m=\overline{\langle f \rangle}$ and $V_D$~\cite{tuyuhai,yann}. In the inset, the black curves show the approximate distribution in Eq~\eqref{eq:pexp}.

In Fig~\ref{fig:Fig2}B,D the Langevin trajectories were generated using Euler's method to integrate Eq~\eqref{eq:Langevin}.

In Fig~\ref{fig:Fig3}C,F,I the $\tau_E$ calculation considered receptor saturation as well as the varying gradient length scales, with $C$ and $L$ evaluated at mean positions. Note this is not the average $\tau_E$ over the population.

\section*{Acknowledgments}

We thank D Clark, Y Dufour, N Frankel, X Fu, S Kato, N Olsman, DC Vural, and A Waite for discussions. This work was supported by the HPC facilities operated by, and the staff of, the Yale Center for Research Computing. JL received support from the Natural Sciences and Engineering Research Council of Canada (NSERC) Postgraduate Scholarships-Doctoral Program (http://www.nserc-crsng.gc.ca/Students-Etudiants/PG-CS/BellandPostgrad-BelletSuperieures\_eng.asp) PGSD2-471587-2015. TE received support from the National Institute of General Medical Sciences (www.nigms.nih.gov) grant 4R01GM106189-04. TE and SWZ received support from the Allen Distinguished Investigator Program (grant 11562) through the Paul G. Allen Frontiers Group (www.pgafamilyfoundation.org/programs/investigators-fellows). The funders had no role in study design, data collection and analysis, decision to publish, or preparation of the manuscript.

%\nolinenumbers

\section*{Supporting Information}

% Include only the SI item label in the paragraph heading. Use the \nameref{label} command to cite SI items in the text.
We provide the Supporting Information in a single file with the following table of contents:

\paragraph*{S1 Appendix.}
\label{S1}
{\bf Agent-based Models and Numerical Methods}

\paragraph*{S1 Fig.}
\label{S1_Fig}
{\bf Robustness of results.}

\paragraph*{S2 Fig.}
\label{S2_Fig}
{\bf Effect of changing $\tau_D$.}

\paragraph*{\textcolor{black}{S3 Fig}.}
\label{S3_Fig}
{\textcolor{black}{\bf Enhanced chemotaxis with signaling noise.}}

\paragraph*{S1 Table.}
\label{Sparam}
{\bf Parameter values used in agent-based simulations.}

\paragraph*{\textcolor{black}{S1 Movie}.}
\label{S1_Movie}
\textcolor{black}{{\bf Movie of the $(r,v)$ phase space trajectories shown in Fig~\ref{fig:Fig2}.}}

\setboolean{@twoside}{false}


\section*{Supplementary material}

# S1 Appendix. Agent-based Models and Numerical Methods

## Model of the chemotaxis pathway

We used a standard model of *E. coli* chemotaxis [1–5] in which the cell relays information from the external environment to the flagellar motor through a signaling cascade triggered by the binding of ligand to transmembrane chemoreceptors (parameters in Table S1). The receptors form cooperative clusters, the activity of which is described by a two-state model where the activity $a$ is determined by the free energy difference $F$ (in units of $k_B T$) between the active and inactive states

$$a = \frac{1}{1 + \mathrm{e}^F}. \tag{S1}$$

Two terms, $F = F_m(m) + F_C(C)$, contribute to the free energy. The former depends on the methylation level $m$ of the receptor

$$F_m(m) = \epsilon_0 + \epsilon_1 m, \tag{S2}$$

with constants $\epsilon_0$ and $\epsilon_1$, while the latter depends on the ligand concentration $C$

$$F_C(C) = N^{Rec} \ln \frac{1 + C/K_i}{1 + C/K_a}. \tag{S3}$$

where $K_a$ and $K_i$ are the ligand dissociation constants of the active and inactive states and $N^{Rec}$ is the degree of cooperativity of the cluster, i.e. the average size of the subclusters that switch as all-or-none units within the cluster of receptors.

As the external environment signals changes in $F_C(C)$, the receptor adapts via methylation and demethylation to control $F_m(m)$, trying to maintain an activity level $a_0$ independent of the environment. The methylation kinetics is described by Eq (4) of [3]

$$\frac{\mathrm{d}m}{\mathrm{d}t} = V_R \frac{1 - a}{K_R + 1 - a} - V_B(a) \frac{a}{K_B + a}, \tag{S4}$$

where $V_R$ and $V_B(a) = V_B(0) \left( 1 + \theta(a - a_B) \frac{a - a_B}{1 - a_B} r_B \right)$ ($\theta$ is the Heaviside function and $a_B = 0.74$ and $r_B = 4.0$) are the methylation/demethylation rates of the proteins CheR and CheB, and $K_R$ and $K_B$ are their dissociation constants respectively. Because $\epsilon_1 < 0$ (Table S1), Eqs (S1)–(S4) show that methylation and demethylation reactions tends to maintain the system at a constant activity level $a_0$.

Assuming the signal transduction is fast compared to adaptation kinetics, the receptor activity determines the concentration of the response regulator CheY-P, $Y(a) = \alpha a$ with $\alpha$ constant. CheY-P then binds to the flagellar motor complex, modulating its switching between rotating clockwise (CW) or counterclockwise (CCW), which are Poisson processes with rates from CCW to CW as $\lambda_{CCW}(Y)$ and from CW to CCW as $\lambda_{CW}(Y)$

$$\lambda_{CCW,CW}(Y) = \omega \exp\left(\mp G(Y)\right), \quad \text{with} \quad G(Y) = \frac{\epsilon_2}{4} - \frac{\epsilon_3}{2} \frac{1}{1 + K/Y}, \tag{S5}$$

where $\omega$, $\epsilon_2$, $\epsilon_3$, and $K$ are constants.

The motor state of each flagellum determines its conformation to be in one of normal, curly-1, and semi-coiled. The flagellar conformations in turn determine the cell's motility state (run or tumble). For a single-flagellum cell, the motor states CCW and CW corresponds to the motility states run and tumble, and we can write $\lambda_R = \lambda_{CCW}$ and $\lambda_T = \lambda_{CW}$ as the switching rates from run to tumble and from tumble to run, respectively [4].

## Analytical approximations

In this section we show how the standard bacterial chemotaxis model just described maps onto the minimal model of run-and-tumble navigation used for analytical derivations in the main text. Doing so requires

making simplifying assumptions. In Fig S1D,F we verify the validity of these approximations by running agent-based simulations with and without such simplifications.

Following [4] methylation-demethylation kinetics are modeled as a linear relaxation around the adapted level of methylation $m_0(C)$ that maintains $a_0$. From Eqs (S1)–(S4) we get

$$\frac{\mathrm{d}m}{\mathrm{d}t} = -\frac{m - m_0(C)}{t_M},$$

(S6)

where $t_M$ is the adaptation time and

$$m_0(C) = \frac{1}{\epsilon_1}\left(F_0 - N^{Rec}\ln\frac{1 + C/K_i}{1 + C/K_a} - \epsilon_0\right), \quad \text{with} \quad F_0 = -\ln\left(1/a_0 - 1\right).$$

(S7)

so that when $m = m_0(C)$ it follows that $F = F_0$ and $a = a_0$. Substituting in Eq (S6) we obtain:

$$\frac{\mathrm{d}F}{\mathrm{d}t} = -\frac{F - F_0}{t_M} + N^{Rec}\frac{\mathrm{d}}{\mathrm{d}t}\ln\frac{1 + C/K_i}{1 + C/K_a} = -\frac{F - F_0}{t_M} + N^{Rec}\left(\frac{1}{K_i/C + 1} - \frac{1}{K_a/C + 1}\right)\frac{\mathrm{d}}{\mathrm{d}t}\ln C,$$

(S8)

where $\mathrm{d}/\mathrm{d}t = \partial/\partial t + \dot{\mathbf{X}}\cdot\nabla$ is the material derivative along the path of the cell. Comparing this equation with Eq (1) in the main text, we define the perceived signal as the log-concentration

$$\phi = \ln\frac{C}{K_i},$$

(S9)

while the receptor gain is concentration-dependent

$$N = N^{Rec}\left(\frac{1}{K_i/C + 1} - \frac{1}{K_a/C + 1}\right).$$

(S10)

The concentration-dependent factor is always less than 1, and as $C$ increases this factor contributes to a smaller gain. Within the range of sensitivity of the receptor, i.e. when $K_i \ll C \ll K_a$, this factor is close to 1 and we obtain the log-sensing approximation. In this case, Eq (S7) can be further simplified to

$$m_0(C) \approx \frac{1}{\epsilon_1}\left(F_0 - N^{Rec}\ln\frac{C}{K_i} - \epsilon_0\right), \quad \text{with} \quad F_0 = -\ln\left(1/a_0 - 1\right).$$

(S11)

Given constant Poisson switching rates $\lambda_R$ and $\lambda_T$, the probability to be running is determined by $r = \lambda_T/\left(\lambda_R + \lambda_T\right)$. From Eq (S5) and the definition $Y = \alpha a = \alpha/\left(1 + e^F\right)$ we can write

$$r(F) = \frac{1}{1 + \frac{\lambda_R}{\lambda_T}} = \frac{1}{1 + \exp\left(-\frac{\epsilon_2}{2} + \epsilon_3\frac{1}{1 + \frac{K}{\alpha}\left(1 + e^F\right)}\right)}.$$

(S12)

Thus the probability to run $r$ is a monotonically increasing function of the free energy $F$. Its shape is almost identical to the standard sigmoidal function $1/\left(1 + \exp\left(-H\left(F - \delta\right)\right)\right)$ with a scaling $H$ and shift $\delta$. Linearly expanding both expressions and matching zeroth and first order we obtain:

$$\delta = -\ln\left(\left(2\epsilon_3/\epsilon_2 - 1\right)\alpha/K - 1\right)$$
$$H = \epsilon_3 e^{-\delta}\left(\frac{\epsilon_2}{2\epsilon_3}\right)^2\frac{K}{\alpha}.$$

(S13)

For the parameters chosen we have $\delta \approx 0$ (Table S1). Therefore, $r = \lambda_T/\left(\lambda_R + \lambda_T\right) \approx 1/\left(1 + \exp\left(-HF\right)\right)$.

## Agent-based simulations

The agent-based chemotaxis simulations were performed using Euler's method as described previously [1, 4, 6, 7]. At each time, each cell moves forward or stays in place according to its motility state (run or tumble), which also determines whether its direction changes with rotational diffusion coefficients $D_R$ or

$D_T$. Once the position, direction and local ligand concentrations are updated, the adaptation equation is integrated and the free energy difference and activity of the receptors is updated. This in turn determines the switching rates according to Eq (S5) and $Y(t) = \alpha a(t)$, and the motor state is changed if a random number drawn exceeds the probability to switch over the time step. After that the motor state determines the flagellum state and subsequently the cell motility state with rules and parameters as in [4], completing one time step.

We used three different types of agent-based simulation:

1. The full nonlinear simulation integrated equations Eqs (S1)–(S5) and served as our reference model of bacterial chemotaxis against which we tested our approximations (Fig S1F).

2. The second model [1, 4] used linear adaptation kinetics replacing Eq (S4) with Eqs (S6)–(S7). This model was used to generate Fig 3 as well as the scatter plot in Fig 1A. Note that for the latter, the values of the parameters $t_M$, $f_0$, $v_0$, $D_R$ and $D_T$ where chosen according to random distributions to match the phenotypic heterogeneity measured in wild type population of *E. coli* (see below).

3. The third model was the same as the second one with the added simplification that the cell was assumed to be perfectly log sensing: Eq (S11) was used instead of Eq (S7). This model was used to generate the heat map in Fig 1A.

## Simulation setup

In Fig 1A heat map, model 3 was used to generate $10^4$ sample trajectories that were each 200 seconds long — sufficient for the cells to reach steady state — for each parameter set $\tau_E$ and $\tau_{D0}$. We used constant run speed $v_0 = 20 \mu m/s$, memory length $t_M = 10\ s$, and adapted probability to run $r_0 = 0.8$ for all cells and all parameter sets. The gradient length scale $L$ and rotational diffusion coefficients $D_R$ and $D_T$ were then determined by each timescale ratios. We started the cells at the bottom of an exponential gradient where the initial concentration is $C_i = 0.1\ mM$.

In Fig 1A scatter plot, model 2 together with the noisy gene expression model and parameter values as in [7] was used to generate 16,000 cells with different run speeds $v_0$, memory timescales $t_M$ and run probabilities $r_0$. We varied $D_R$ and $D_T$ across cells to account for variations in the cell lengths and their effect on the rotational diffusion coefficients. Noting that the coefficient of variation (CV) of the *E. coli* cell lengths is $\sim 15\%$ [8], and that the rotational diffusion coefficient of an ellipsoid with major axis (length $l$) parallel to the direction of motion has $D_R \sim l^{-3}$ [9], we estimated the CV of $D_R$ to be about 3 times that of the cell lengths, or $\sim 50\%$. Therefore, we sampled $D_R$ from a log-normal distribution (to make sure $D_R > 0$) with mean $0.062\ s^{-1}$ [9] and standard deviation $0.03\ s^{-1}$. Assuming that $D_T$ is affected similarly by the cell length, we fixed the ratio $D_T/D_R \approx 37$ across all cells. We simulated the cells in a quasi-linear gradient (fit from experimental data as described in [7]) of methyl-aspartate that varies from 0 to 1 $mM$ over 10 $mm$. Near the bottom of the gradient (at $x = 1\ mm$) we calculated $L = 1500\ \mu m$ and near the top of the gradient (at $x = 9\ mm$) we found $L = 4800\ \mu m$.

In Fig 3 model 2 was used to generate $10^4$ sample trajectories for each parameter set $\tau_{D0} = 1$ and $\tau_E = 0.1, 1, 3$, for a total time of 650 seconds in the exponential gradient, 2000 seconds in the linear gradient, and 3000 seconds for the localized source. We used constant run speed $v_0 = 20 \mu m/s$ and adapted probability to run $r_0 = 0.8$ for all cells and all parameter sets. In all cases the initial length scale where the cells start was $L_i = 1000\ \mu m$, from which we determined the cell memory $t_M$ and the diffusion coefficients $D_R$ and $D_T$ using the timescale ratios.

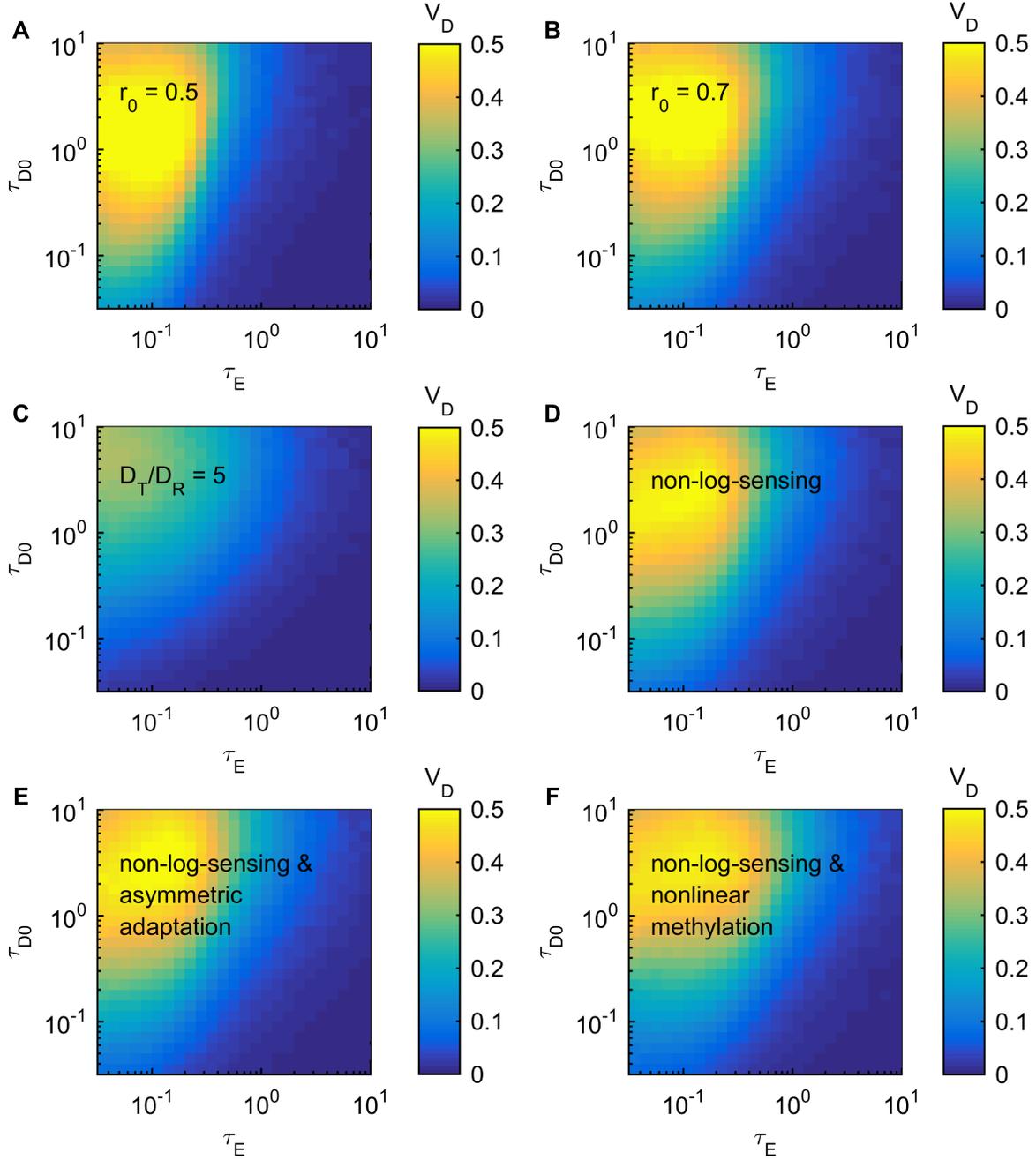

**Fig S1. Robustness of results.**
(A) Same as Fig 1A (where $r_0 = 0.8$, Table S1) except with $r_0 = 0.5$. (B) $r_0 = 0.7$. (C) Same as Fig 1A (where $D_T/D_R \approx 37$, Table S1) except with $D_T/D_R = 5$. (D) Same as Fig 1A except without assuming receptors in log-sensing range, i.e. Eq (S7) was used rather than Eq (S11). (E) Same as D, but additionally implements adaptation asymmetry, where the adaptation rate $t_M^{-1}$ in equation Eq (S6) depends on $m$ [10]. Here the adaptation is 3 times faster when $m > m(C)$ than when $m < m(C)$, and $t_M$ is defined as the time scale when $m < m(C)$. (F) Same as D but with nonlinear adaptation rate Eq (S4). Values for the parameters are: $a_B = 0.74$, $r_B = 4.0$, $K_R = 0.32$, $K_B = 0.30$, and $V_R$ and $V_B(0)$ chosen to ensure $\frac{dm}{dt} = 0$ when $a = a_0$ and the adaptation time is $t_M$ when linearized.

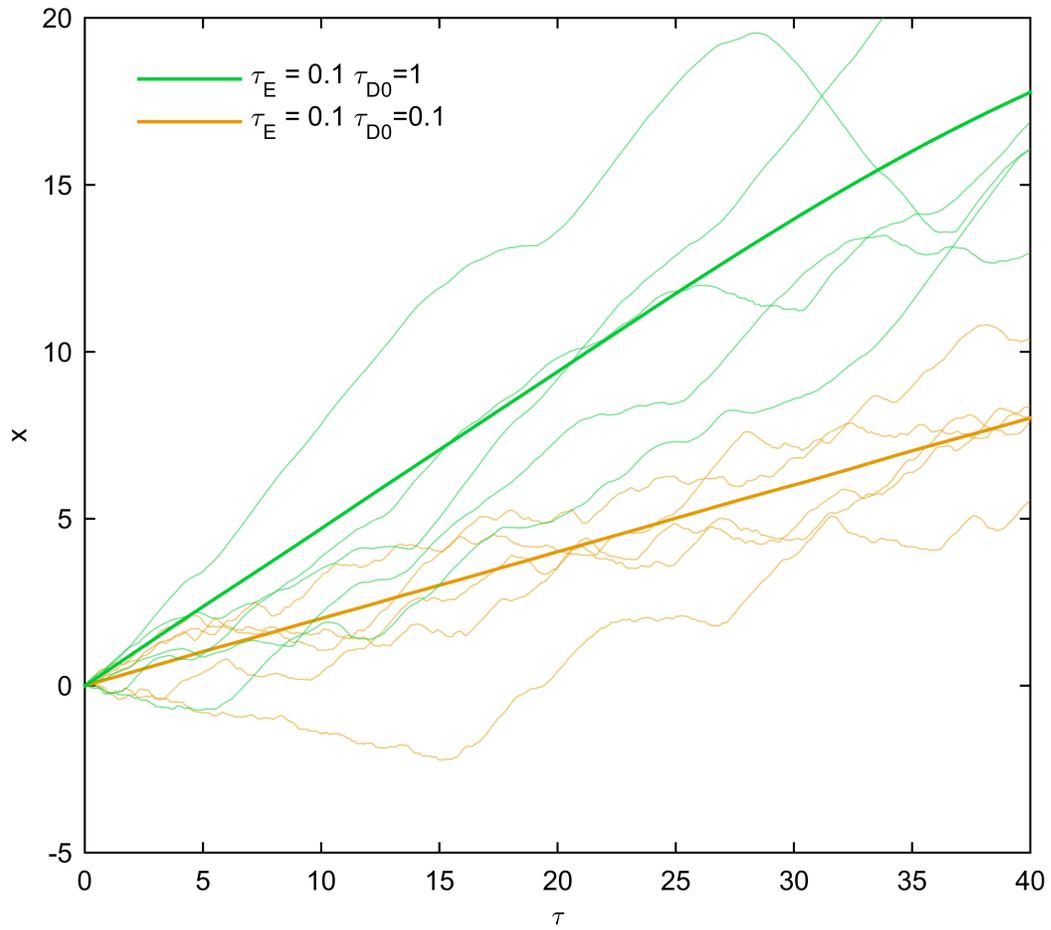

**Fig S2. Effect of changing $\tau_D$.**
5 sample trajectories (thin solid curves) and the mean over $10^4$ sample trajectories (thick lines) of non-dimensionalized position $x = X/(v_0 t_M)$ as a function of time $\tau = t/t_M$. Colors correspond to cells with matching (green $\tau_{D0} = 1$) and non-matching (orange $\tau_{D0} = 0.1$) reorientation as in Fig 1 (same methods).

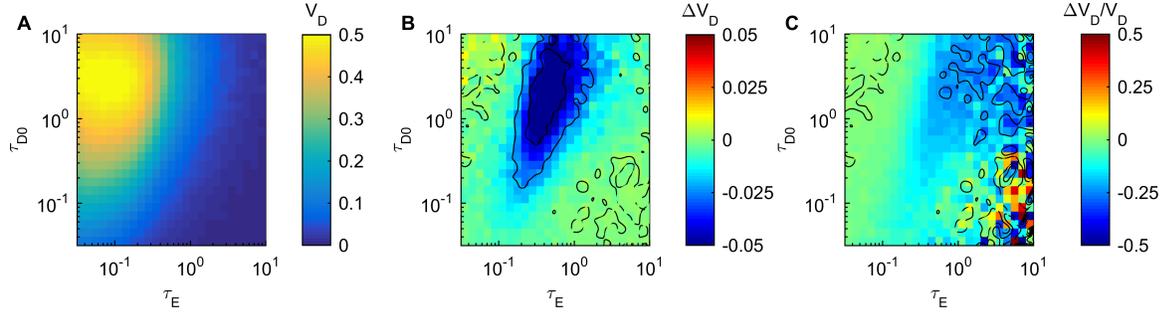

**Fig S3. Enhanced chemotaxis with signaling noise.**
(A) Same as Fig 1A but adding a signaling noise term $\sigma_m \sqrt{2/t_M} \Gamma(t)$ in Eq (S6) where $\Gamma(t)$ is the standard Wiener process (see Eq [4] in [4]). From Eqs (S1)-(S2) and $Y = \alpha a$ we obtain $\mathrm{d}Y/\mathrm{d}m = Y(1-a)\epsilon_1$. Then plugging in $\sigma_Y/Y = 0.1$ [4,11], $a \approx 0.5$ and $\epsilon_1 = -1$, we obtain $\sigma_m = 0.2$. (B) The absolute difference in drift speed between A (with signaling noise) and Fig 1A (without signaling noise), $\Delta V_D = V_D|_{noise} - V_D|_{no\ noise}$, shows how signaling noise can either enhance or reduce the drift speed depending on $(\tau_E, \tau_D)$. Note the colorbar is different. Black contour lines show level sets of $\Delta V_D$ at the colorbar ticks (-0.05, -0.025, 0, 0.025, and 0.05). (C) The relative difference in drift speed (B divided by the drift speed without signaling noise). Again, the color scale is different and black contour lines show level sets at the colorbar ticks.

**Table S1. Parameter values used in agent-based simulations.**

| Fixed Parameters | | | |
|---|---|---|---|
| Name | Definition | Value | References and Explanations |
| $\epsilon_0$ | Eq (S2) | 6 | Shimizu *et al.* 2010 [3] |
| $\epsilon_1$ | Eq (S2) | $-1$ | Shimizu *et al.* 2010 [3] |
| $N^{Rec}$ | Eq (S3) | 6 | Shimizu *et al.* 2010 [3] |
| $K_i$ | Eq (S3) | $0.0182\ mM$ | Shimizu *et al.* 2010 [3] |
| $K_a$ | Eq (S3) | $3\ mM$ | Shimizu *et al.* 2010 [3] |
| $\alpha$ | $Y = \alpha a$ | 6 | Sneddon *et al.* 2012 [4] |
| $K$ | Eq (S5) | $3.06\ mM$ | Sneddon *et al.* 2012 [4] |
| $\epsilon_2$ | Eq (S5) | 40 | Sneddon *et al.* 2012 [4] |
| $\epsilon_3$ | Eq (S5) | 40 | Sneddon *et al.* 2012 [4] |
| $\omega$ | Eq (S5) | $1.3\ s^{-1}$ | Sneddon *et al.* 2012 [4] |
| $\delta$ | Eq (S13) | $-0.04$ | Equation Eq (S13) |
| $H$ | Eq (S13) | 4.9 | Equation Eq (S13) |
| **Parameters in Fig 1A heatmap** | | | |
| Name | Definition | Value | References and Explanations |
| $v_0$ | Maximal run speed | $20\ \mu m/s$ | Sneddon *et al.* 2012 [4] |
| $r_0$ | $r(f = f_0)$ | 0.8 | This study |
| $t_M$ | Memory | $10\ s$ | Dufour *et al.* 2014 [1] |
| $C_i$ | Initial concentraion | $0.1\ mM$ | This study |
| $D_R$ | Rotational diffusion (run) | $0.00061 - 0.19\ s^{-1}$ | Let $\tau_{D0}$ vary from $10^{-1.5}$ to $10^1$ |
| $D_T$ | Rotational diffusion (tumble) | $0.023 - 7.1\ s^{-1}$ | Keep $D_T/D_R \approx 37$ [9,12] |
| $L$ | Gradient length scale | $186 - 58800\ \mu m$ | Let $\tau_E$ vary from $10^{-1.5}$ to $10^1$ |
| **Parameters in Fig 1A scatter plot that are different from [7]** | | | |
| Name | Definition | Value | References and Explanations |
| $D_R$ | Rotational diffusion (run) | sampled from log-normal: mean $0.062\ s^{-1}$ std $0.03\ s^{-1}$ | Account for variations in cell lengths [8,9] |
| $D_T$ | Rotational diffusion (tumble) | $37\ D_R$ | Keep $D_T/D_R \approx 37$ [9,12] |
| $L$ | Gradient length scale | $1500\ \mu m$ and $4800\ \mu m$ | $1/||\nabla \ln C||$ |
| **Parameters in Fig 3** | | | |
| Name | Definition | Value | Reference and Explanations |
| $v_0$ | Maximal run speed | $20\ \mu m/s$ | Sneddon *et al.* 2012 [4] |
| $r_0$ | $r(f = f_0)$ | 0.8 | This study |
| $L_i$ | Initial gradient length scale | $1000\ \mu m$ | This study |
| $C_0$ | Exponential gradient source | $10\ mM$ | This study |
| $L_0$ | Length scale of $C_0$ source | $1000\ \mu m$ | This study |
| $C_1$ | Linear gradient source | $1\ mM$ | This study |
| $a_1$ | Slope of $C_1$ source | $0.0001\ \mu M/\mu m$ | This study |
| $C_2$ | Localized source | $1\ mM$ | This study |
| $R_0$ | Size of $C_2$ source | $100\ \mu m$ | This study |
| $t_M$ | Memory | $0.54 - 17\ s$ | Let initial $\tau_E$ vary from 3 to 0.1 |
| $D_R$ | Rotational diffusion (run) | $0.0036 - 0.11\ s^{-1}$ | Keep $\tau_D = 1$ fixed |
| $D_T$ | Rotational diffusion (tumble) | $0.13 - 4.2\ s^{-1}$ | Keep $D_T/D_R \approx 37$ [9,12] |




\begin{thebibliography}{70}

\bibitem{bergbrown}
Berg HC, Brown, DA.
\newblock {{C}hemotaxis in \textit{{E}scherichia coli} analysed by three-dimensional tracking}.
\newblock Nature. 1972 Oct;239:500--504.

\bibitem{marineMicrobes}
Stocker R.
\newblock {{M}arine microbes see a sea of gradients}.
\newblock Science. 2012 Nov;338:628--633.
	
\bibitem{albrecht-bargmann}
Albrecht DR, Bargmann CI.
\newblock {{H}igh-content behavioral analysis of \textit{{C}aenorhabditis elegans} in precise spatiotemporal chemical environments}.
\newblock Nat Methods. 2011 Jul;8(7):599--606.

\bibitem{drosophilaChemotaxis}
Gomez-Marin A, Stephens GJ, Louis M.
\newblock {{A}ctive sampling and decision making in \textit{{D}rosophila} chemotaxis}.
\newblock Nat Commun. 2011 Mar;2:441.

\bibitem{swarmRobotics}
Taylor-King JP, Franz B, Yates CA, Erban R.
\newblock {{M}athematical modelling of turning delays in swarming robots}.
\newblock IMA J Appl Math. 2015 Mar;80:1454--1474.

\bibitem{doyle}
Yi TM, Huang Y, Simon MI, Doyle J.
\newblock {{R}obust perfect adaptation in bacterial chemotaxis through integral feedback control}.
\newblock Proc Natl Acad Sci U S A. 2000 Apr;97(9):4649--4653.

\bibitem{trefethen}
Trefethen LN, Trefethen AE, Reddy SC, Driscoll TA.
\newblock {{H}ydrodynamic {S}tability {W}ithout {E}igenvalues}.
\newblock Science. 1993 Jul;261:578--584.

\bibitem{nonmodal}
Schmid PJ.
\newblock {{N}onmodal {S}tability {T}heory}.
\newblock Annu Rev Fluid Mech. 2007 Jan;39:129--162.

\bibitem{KS}
Keller EF, Segel LA.
\newblock {{T}raveling bands of chemotactic bacteria: a theoretical analysis}.
\newblock J Theor Biol. 1971;30:235--248.

\bibitem{schnitzer}
Schnitzer MJ.
\newblock {{T}heory of continuum random walks and application to chemotaxis}.
\newblock Phys Rev E. 1993 Oct;48(4):2553--2568.

\bibitem{vergassola}
Celani A, Vergassola M.
\newblock {{B}acterial strategies for chemotaxis response}.
\newblock Proc Natl Acad Sci U S A. 2010 Jan;107(4):1391--1396.

\bibitem{tuyuhai}
Si G, Wu T, Ouyang Q, Tu Y.
\newblock {{P}athway-based mean-field model for \textit{{E}scherichia coli} chemotaxis}.
\newblock Phys Rev Lett. 2012 Jul;109(4):048101.

\bibitem{yann}
Dufour YS, Fu X, Hernandez-Nunez L, Emonet T.
\newblock {{L}imits of feedback control in bacterial chemotaxis}.
\newblock PLoS Comput Biol. 2014 Jun;10:e1003694.

\bibitem{shimizu-tu-berg}
Shimizu TS, Tu Y, Berg HW.
\newblock {{A} modular gradient-sensing network for chemotaxis in \textit{{E}scherichia coli} revealed by responses to time-varying stimuli}.
\newblock Mol Syst Biol. 2010 Jun;6:382--395.

\bibitem{nick}
Frankel NW, Pontius W, Dufour YS, Long J, Hernandez-Nunez L, \textit{et al}.
\newblock {{A}daptability of non-genetic diversity in bacterial chemotaxis}.
\newblock eLife. 2014 Oct;10.7554/eLife.03526.

\bibitem{frequencyDependent}
Zhu X, Si G, Deng N, Ouyang Q, Wu T, \textit{et al}.
\newblock {{F}requency-dependent \textit{{E}scherichia coli} chemotactic behavior}.
\newblock Phys Rev Lett. 2012 Mar;108(12):128101.

\bibitem{xue-yang}
Xue C, Yang X.
\newblock {{M}oment-flux models for bacterial chemotaxis in large signal gradients}.
\newblock J Math Biol. 2016 Feb;doi:10.1007/s00285-016-0981-9:1--24.

\bibitem{tuReview}
Tu Y.
\newblock {{Q}uantitative modeling of bacterial chemotaxis signal amplification and accurate adaptation}.
\newblock Annu Rev Biophys. 2013 Feb;42:337--359.

\bibitem{saragosti2012}
Saragosti J, Silberzan P, Buguin A.
\newblock {{M}odeling \textit{{E}. coli} tumbles by rotational diffusion. Implications for chemotaxis}.
\newblock PLoS ONE. 2012 Apr;7(4):e35412.

\bibitem{michael}
Sneddon MW, Pontius W, Emonet T.
\newblock {{S}tochastic coordination of multisple actuators reduce latency and improves chemotactic response in bacteria}.
\newblock Proc Natl Acad Sci U S A. 2012 Jan;109(2):805--810.

\bibitem{tenwolde}
Flores M, Shimizu TS, ten Wolde PR, Tostevin F.
\newblock {{S}ignaling {N}oise {E}nhances {C}hemotactic {D}rift of \textit{{E}. coli}}.
\newblock Phys Rev Lett. 2012 Oct;109:148101.

\bibitem{adam}
Waite AJ, Frankel NW, Dufour YS, Johnston JF, Long J, Emonet T.
\newblock {{N}on-genetic diversity modulates population performance}.
\newblock Mol Syst Biol. 2016 accepted.

\bibitem{damon}
Clark DA, Grant LC.
\newblock {{T}he bacterial chemotactic response reflects a compromise between transient and steady-state behavior}.
\newblock Proc Natl Acad Sci U S A. 2005 Jun;102(26):9150--9155.

\bibitem{rava}
Kafri Y, da Silveira RA.
\newblock {{S}teady-state chemotaxis in \textit{{E}scherichia coli}}.
\newblock Phys Rev Lett. 2008 Jun;100(23):238101.

\bibitem{yannExperimental}
Dufour YS, Gillet S, Frankel NW, Weibel DB, Emonet T.
\newblock {{D}irect correlation between motile behaviors and protein abundance in single cells}.
\newblock PLoS Comput Biol. 2016 Sep;12(9):e1005041.

\bibitem{spudichKoshland1976}
Spudich JL, Koshland DE Jr.
\newblock {{N}on-genetic individuality: chance in the single cell}.
\newblock Nature. 1976 Aug;262:467--471.

\bibitem{neuron}
Murphy BK, Miller KD.
\newblock {{B}alanced amplification: a new mechanism of selective amplification of neural activity patterns}.
\newblock Neuron. 2009 Feb;61:635--648.

\bibitem{NNampli}
Hennequin G, Vogels TP, Gerstner W.
\newblock {{N}on-normal amplification in random balanced neural networks}.
\newblock Phys Rev E. 2012 Jul;86(1):011909.

\bibitem{colin}
Colin R, Zhang R, Wilson LG.
\newblock {{F}ast, hight-throughput measurement of collective behavior in a bacterial population}.
\newblock J R Soc Interface. 2014 Jul;11:20140486.

\bibitem{mello-tu}
Mello BA, Tu Y.
\newblock {{Q}uantitative modeling of sensitivity in bacterial chemotaxis: {T}he role of coupling among different chemoreceptor species}.
\newblock Proc Natl Acad Sci U S A. 2003 Jul; 100(14):8223--8228.

\bibitem{sourjik-berg}
Sourjik V, Berg HC.
\newblock {{F}unctional interactions between receptors in bacterial chemotaxis}.
\newblock Nature. 2004 Mar;428:437--441.

\bibitem{keymer-endres-wingreen}
Keymer JE, Endres RG, Skoge M, Meir Y, Wingreen NS.
\newblock {{C}hemosensing in \textit{{E}scherichia coli}: {T}wo regimes of two-state receptors}.
\newblock Proc Natl Acad Sci U S A. 2006 Feb;103(6):1786--1791.

\bibitem{hansen-endres-wingreen}
Hansen CH, Endres RG, Wingreen NS.
\newblock {{C}hemotaxis in \textit{{E}scherichia coli}: {A} {M}olecular {M}odel for {R}obust {P}recise {A}daptation}.
\newblock PLoS Comput Biol. 2008 Jan;4(1):e1.

\bibitem{blackburn}
Blackburn N, Fenchel T, Mitchell J.
\newblock {{M}icroscale {N}utrient {P}atches in {P}lanktonic {H}abitats {S}hown by {C}hemotactic {B}acteria}.
\newblock Science. 1989 Dec;282:2254--2256.

\bibitem{predictEnv}
Clausznitzer D, Micali G, Neumann S, Sourjik V, Endres RG.
\newblock {{P}redicting {C}hemical {E}nvironments of {B}acteria from {R}eceptor {S}ignaling}.
\newblock PLoS Comput Biol. 2014 Oct;10(10):e1003870.

\bibitem{physicalLimits}
Hein AM, Brumley DR, Carrara F, Stocker R, Levin SA.
\newblock {{P}hysical limits on bacterial navigation in dynamic environments}.
\newblock J R Soc Interface. 2016 Jan;13:20150844.

\bibitem{tindall}
Edgington MP, Tindall MJ.
\newblock {{U}nderstanding the link between single cell and population scale responses of \textit{{E}scherichia coli} in differing ligand gradients}.
\newblock Comput Struct Biotechnol J. 2015 Oct;13:528--538.

\bibitem{elowitz}
Elowitz MB, Levine AJ, Siggia ED, Swain PS.
\newblock {{S}tochastic {G}ene {E}xpression in a {S}ingle {C}ell}.
\newblock Science. 2002 Aug;297:1183--1186.

\bibitem{kussell}
Kussell E, Leibler S.
\newblock {{P}henotypic {D}iversity, {P}opulation {G}rowth, and {I}nformation in {F}luctuating {E}nvironments}.
\newblock Science. 2005 Sep;309:2075--2078.

\bibitem{oudenaarden}
Acar M, Mettetal JT, van Oudenaarden A.
\newblock {{S}tochastic switching as a survival strategy in fluctuating environments}.
\newblock Nat Genet. 2008 Apr;40(4):471--475.

\bibitem{ackermann}
Ackermann M.
\newblock {{A} functional perspective on phenotypic heterogeneity in microorganisms}.
\newblock Nat Rev Microbiol. 2015 Aug;13:497--508.

\bibitem{asym-sourjik-endres}
Clausznitzer D, Oleksiuk O, Løvdok L, Sourjik V, Endres RG.
\newblock {{C}hemotactic {R}esponse and {A}daptation {D}ynamics in \textit{{E}scherichia coli}}.
\newblock PLoS Comput Biol. 2010 May;6(5):e1000784.

\bibitem{bergCell}
Block SM, Segall JE, Berg HC.
\newblock {{I}mpulse {R}esponses in {B}acterial {C}hemotaxis}.
\newblock Cell. 1982 Nov;31:215--226.

\bibitem{vergassolaRace}
Wong-Ng J, Melbinger A, Celani A, Vergassola.
\newblock {{T}he {R}ole of {A}daptation in {B}acterial {S}peed {R}aces}.
\newblock PLoS Comput Biol. 2016 Jun;12(6):e1004974.

\bibitem{stochastic}
Gardiner CW.
\newblock Handbook of stochastic methods for physics, chemistry and the natural sciences.
\newblock Berlin: Springer Berlin Heidelberg; 2004.

\bibitem{geometry}
Jost J.
\newblock Riemannian Geometry and Geometric Analysis.
\newblock Berlin: Springer Berlin Heidelberg; 2011.

\bibitem{Gegenbauer}
Abramowitz M, Stegun I.
\newblock Handbook of Mathematical Functions with Formulas, Graphs, and Mathematical Tables.
\newblock Mineola: Dover Publications; 1964.

\bibitem{erban-othmer2004}
Erban R, Othmer HG.
\newblock {{F}rom individual to collective behavior in bacteria chemotaxis}.
\newblock SIAM J Appl Math. 2004 Dec;65(2):361--391.

\bibitem{franz}
Franz B, Xue C, Painter KJ, Erban R.
\newblock {{T}ravelling {W}aves in {H}ybrid {C}hemotaxis {M}odels}
\newblock Bull Math Biol. 2014 Feb;76(2):377--400.

\end{thebibliography}
\end{document}